\documentclass[aps,prd,twocolumn,groupedaddress,nofootinbib,longbibliography]{revtex4-1}

\usepackage{amssymb}
\usepackage{amsmath}
\usepackage[linktocpage=true]{hyperref}

\begin{document}

\title{Spinning particles in vacuum space-times of different curvature types}

\author{O. Semer\'ak}
\email[]{oldrich.semerak@mff.cuni.cz}

\author{M. \v{S}r\'amek}
\email[]{kemrash@seznam.cz}

\affiliation{Institute of Theoretical Physics, Faculty of Mathematics and Physics, Charles University in Prague, Czech Republic}

\date{\today}

\begin{abstract}
We consider the motion of spinning test particles with non-zero rest mass in the ``pole-dipole" approximation, as described by the Mathisson--Papapetrou--Dixon (MPD) equations, and examine its properties in dependence on the spin supplementary condition added to close the system. In order to more understand the spin-curvature interaction, the MPD equation of motion is decomposed in the orthonormal tetrad whose time vector is given by the four-velocity $V^\mu$ chosen to fix the spin condition (the ``reference observer") and the first spatial vector by the corresponding spin $s^\mu$; such projections do not contain the Weyl scalars $\Psi_0$ and $\Psi_4$ obtained in the associated Newman--Penrose (NP) null tetrad. One natural option how to choose the remaining two spatial basis vectors is shown to follow ``intrinsically" whenever $V^\mu$ has been chosen; it is realizable if the particle's four-velocity and four-momentum are not parallel. In order to see how the problem depends on the algebraic type of curvature, one first identifies the first vector of the NP tetrad $k^\mu$ with the highest-multiplicity principal null direction of the Weyl tensor, and then sets $V^\mu$ so that $k^\mu$ belong to the spin-bivector eigen-plane. In space-times of any algebraic type but III, it is known to be possible to rotate the tetrads so as to become ``transverse", namely so that $\Psi_1$ and $\Psi_3$ vanish. If the spin-bivector eigen-plane could be made coincide with the real-vector plane of any of such transverse frames, the spinning-particle motion would consequently be fully determined by $\Psi_2$ and the cosmological constant; however, this can be managed in exceptional cases only. Besides focusing on specific Petrov types, we derive several sets of useful relations valid generally and check whether/how the exercise simplifies for some specific types of motion. The particular option of having four-velocity parallel to four-momentum is advocated and a natural resolution of non-uniqueness of the corresponding reference observer $V^\mu$ is suggested.
\end{abstract}

\pacs{04.25.-g}

\maketitle

\section{Introduction}

Curvature of physical space-time is the major prediction of general relativity, so it is of special interest to study processes in which curvature (non-homogeneity of gravitational field) plays a {\em direct} role. Being described by the Riemann tensor, the participation of curvature usually makes the problem difficult, at least in comparison with those involving just metric and/or affine connection. One of such problems is the motion of extended bodies (e.g. \cite{Dixon-15}). Even if the body is treated as non-radiating, test and small (with all the lengths connected with its multipoles much shorter than the space-time curvature radius), the corresponding equations of motion even contain the Riemann tensor together with its derivatives. More explicit studies thus mostly restrict to the ``spinning particle" limit (the ``pole-dipole" approximation) when just monopole (mass) and dipole (rotational angular momentum, spin) are taken into account and the motion is described by the Mathisson--Papapetrou--Dixon (MPD) equations, supplemented by some ``spin condition". The approximation is problematic in highly non-homogeneous fields, mainly due to the disregard for the quadrupole effect, but we adhere to it here.
Of important recent references in the field, let us recommend \cite{SteinhoffP-10,HackmannLOPS-14,Mohseni-15,dAmbrosiSH-15,CostaN-15}.

The curvature properties are often best revealed in a suitable orthonormal tetrad, namely as represented in terms of the ``Weyl scalars" -- independent projections of the Weyl tensor in the attached Newman--Penrose (NP) null tetrad, which can be given a physical interpretation. One can then understand the geometrical/physical effect of the individual curvature terms in dependence on the scalars, and especially discuss the situations when some of the scalars vanish -- the {\it algebraically special} cases. Such studies actually began before the birth of the NP formalism (e.g. \cite{Pirani-56,Pirani-57}) and since then have notably been devoted to geodesic deviation as a universal probe of gravitational-field properties \cite{Szekeres-65,PodolskyS-12} or to interpretation of space-time perturbations \cite{Nolan-04}.\footnote
{Another major curvature-interpretation direction stems from the celebrated analogy between curvature tensors and electromagnetic-field tensor or electromagnetic tidal tensor -- see \cite{Bonnor-95,MaartensB-98,CostaH-08,CostaN-14}, for example.}
Surprisingly, for the spinning-particle problem a similar discussion has been published in the massless case only \cite{BiniCGJ-06}. (The gravito-electromagnetic parallel has been applied to it by \cite{CostaNZ-12}.)

In the present paper, we will consider particles with non-zero mass. We keep the cosmological constant, but restrict to {\em vacuum} space-times (with zero energy-momentum tensor), since otherwise we would also have to incorporate interaction of the particle with matter and/or other physical fields, generally including torques exerted on its spin. This would certainly obscure the effects coming from curvature and its particular algebraic type.
In the following section \ref{MPD-equations}, we first recall the spinning-particle problem, consider its basic properties including the necessity to add a certain ``spin supplementary condition", and write the equations down in terms of the spin vector and Riemann-tensor dual.
In section \ref{MPD-in-tetrad} the equation of motion is expressed in a suitable orthonormal as well as complex null (NP) frame, representing the Weyl-tensor dual in terms of its complex projections $\Psi_0\div\Psi_4$
(details are shifted to appendix \ref{Appendix-A}).
We show that the problem itself provides a tetrad which can be used rather generally (specifically, if the particle's four-velocity and four-momentum are not parallel) and which gives very simple results mainly in connection with Tulczyjew's supplementary condition.

In order to discuss the equation of motion in dependence on the space-time Petrov type, the interpretation frame is then attached to the Weyl-tensor principal directions in section \ref{MPD,Petrov-types}. As an alternative to the above ``intrinsic" tetrad (and the related null one), there arises a generic possibility (only not applicable in algebraic type III) to use ``transverse frames" in which pure-gauge longitudinal-wave effects vanish, but their special turn can  be aligned with the spin structure only in exceptional cases.
The effect of particular spin conditions is checked in section \ref{SSCs} and several special types of motion are discussed in section \ref{special-motions}.
Concluding remarks close the paper with relating the topic to a wider context and with some tips for improvement or alternative view.

Conventions: we use the metric signature ($-$+++) and geometrized units in which $c=1$, $G=1$. Greek indices run 0--3 and summation convention is followed. The dot denotes absolute derivative with respect to the particle's proper time $\tau$ and overbar indicates complex conjugation.
The Riemann tensor is defined according to $V_{\nu;\kappa\lambda}-V_{\nu;\lambda\kappa}={R^\mu}_{\nu\kappa\lambda}V_\mu$ and the Levi-Civita tensor as
\begin{equation}
  \epsilon_{\mu\nu\rho\sigma}=\sqrt{-g}\,[\mu\nu\rho\sigma], \quad
  \epsilon^{\mu\nu\rho\sigma}=-\frac{1}{\sqrt{-g}}\,[\mu\nu\rho\sigma],
\end{equation}
where $g$ is the covariant-metric determinant and $[\mu\nu\rho\sigma]$ is the permutation symbol fixed by $[0123]:=1$.

\section{Mathisson--Papapetrou--Dixon (MPD) equations}
\label{MPD-equations}

In their seminal papers \cite{Mathisson-37,Papapetrou-51,Tulczyjew-59,Dixon-79},
Mathisson, Papapetrou, Tulczyjew and Dixon provided -- following somewhat different approaches -- a full multipole expansion for the extended-body evolution in general relativity. In greater detail, this problem has as yet been studied in the ``pole-dipole" approximation, including computation of generic trajectories in some basic space-times
\cite{Apostolatos-96,Semerak-99,KyrianS-07,LukesGSK-14}
(and of their dynamics over a corresponding phase space
\cite{SuzukiM-99,Hartl-03a,Hartl-03b,KaoCho-05,Singh-05,KoyamaKK-07,Han-08,VerhaarenH-10}),
while a similar study at quadrupole level has only commenced quite recently
\cite{BiniFGO-08,SteinhoffP-12,BiniG-13,BiniG-14a,BiniG-14b}.
If there is no torque exerted on the particle, the pole-dipole Mathisson-Papapetrou-Dixon (MPD) system reads
\begin{eqnarray}
  \dot{p}^\mu &=&
 -\frac{1}{2}\,{R^\mu}_{\nu\kappa\lambda}u^\nu S^{\kappa\lambda},
       \label{Papa-p} \\
  \dot{S}^{\alpha\beta}&=&p^\alpha u^\beta-u^\alpha p^\beta,
       \label{Papa-S}
\end{eqnarray}
where
$u^\mu$ is a tangent to the world-line which represents the particle's history (it is assumed to be time-like and normalized by $u_\sigma u^\sigma=-1$),
$S^{\mu\nu}$ is the particle-spin bivector,
$p^\mu$ denotes the total momentum (assumed to be time-like)
and $m:= -u_\sigma p^\sigma\,(>\!0)$ is the particle's mass in the frame attached to the representative world-line.
Let us also introduce the particle's mass in the frame given by its momentum, ${\cal M}:=\sqrt{-p_\sigma p^\sigma}\,$, and an associated four-velocity, ${\cal U}^\mu:=p^\mu/{\cal M}$.

The second MPD equation demonstrates that the tensor $\dot{S}^{\alpha\beta}$ is simple and time-like, with its blade spanned by $p^\alpha$ and $u^\alpha$. It also implies that the spin-bivector dual
\[{^*\!}S_{\mu\nu}:=\frac{1}{2}\,\epsilon_{\mu\nu\alpha\beta}S^{\alpha\beta}\]
evolves according to
\begin{equation}  \label{dS,dual}
  {^*\!}\dot{S}_{\mu\nu}
     = \frac{1}{2}\,\epsilon_{\mu\nu\alpha\beta}\dot{S}^{\alpha\beta}
     = \epsilon_{\mu\nu\alpha\beta}p^\alpha u^\beta
\end{equation}
since $\dot{\epsilon}_{\mu\nu\alpha\beta}=0$ and dualization thus commutes with covariant differentiation.
(Hence, ${^*\!}\dot{S}_{\mu\nu}$ is simple and space-like, having $p^\mu$ and $u^\mu$ as eigen-directions corresponding to zero eigen-values.)
This in turn yields
\begin{eqnarray}
  \frac{1}{2}\,\dot{S}^{\alpha\beta}\dot{S}_{\alpha\beta}
  =-\frac{1}{2}\,{^*\!}\dot{S}^{\mu\nu}{^*\!}\dot{S}_{\mu\nu}
    &=& {\cal M}^2-m^2 \leq 0\,, \\
  {^*\!}\dot{S}^{\mu\iota}\dot{S}_{\iota\beta} &=& 0 \,. \label{dotdualS*dotS=0}
\end{eqnarray}

Several useful relations can be obtained directly by projections of the second MPD equation (\ref{Papa-S}):
multiplication by $u_\beta$, $p_\beta$, $\dot{u}_\beta$ and $\dot{p}_\beta$ yields, respectively,
\begin{eqnarray}
  p^\alpha &=& m\,u^\alpha-\dot{S}^{\alpha\beta}u_\beta \,, \label{p,u,dSu} \\
  {\cal M}^2 u^\alpha &=& m\,p^\alpha+\dot{S}^{\alpha\beta}p_\beta \,, \label{u,p,dSp} \\
  \dot{m}\,u^\alpha &=& \dot{S}^{\alpha\beta}\dot{u}_\beta \,, \label{dotm.u} \\
  {\cal M}\dot{\cal M}\,u^\alpha &=& \dot{S}^{\alpha\beta}\dot{p}_\beta \,, \label{dotM.u}
\end{eqnarray}
where in the third case the basic property $u_\beta\dot{p}^\beta=0$ of the first MPD equation (\ref{Papa-p}) has been used (``four-force acts perpendicular to four-velocity").
The first two of these equations indicate a momentum-velocity relation ($m$ can be supplied from normalization $u_\alpha u^\alpha=-1$), while the last two lead to mass-evolution formulas
\begin{eqnarray}
  \dot{m} &=& \dot{u}_\alpha u_\beta \dot{S}^{\alpha\beta}
           =  \dot{u}_\alpha \frac{p_\beta}{m}\,\dot{S}^{\alpha\beta}
           =  -\dot{u}_\alpha p^\alpha, \label{dm} \\
   {\cal M}\dot{\cal M} &=& \dot{p}_\alpha u_\beta\dot{S}^{\alpha\beta}
                         =  \dot{p}_\alpha \frac{p_\beta}{m}\,\dot{S}^{\alpha\beta}
                         =  -\dot{p}_\alpha p^\alpha \label{dM}
\end{eqnarray}
(the last expressions follow from the very definitions of $m$ and ${\cal M}$),
but in both cases they provide only partial answer since containing {\em derivative} of $S^{\alpha\beta}$.
It was shown by \cite{ObukhovP-11} that the MPD equations can also be inverted to
\begin{equation}  \label{u-p,Puetzfeld}
  {\cal M}^2 u^\alpha =
   m \left(\tilde{p}^\alpha+
           \frac{2S^{\alpha\beta}R_{\beta\iota\kappa\lambda}\tilde{p}^\iota S^{\kappa\lambda}}
                {4{\cal M}^2+R_{\mu\nu\gamma\delta}S^{\mu\nu}S^{\gamma\delta}}
           \right),
\end{equation}
where\footnote
{Mind the opposite metric signature used in \cite{ObukhovP-11}, resulting in the opposite sign of $S^{\alpha\beta}p_\beta$.}
\begin{equation}  \label{tilde(p)}
  \tilde{p}^\alpha
    := p^\alpha+\frac{1}{m}\,\frac{{\rm D}}{{\rm d}\tau}(S^{\alpha\beta}p_\beta) \,,
\end{equation}
with $\tau$ standing for proper time and $m$ again fixed by $u_\alpha u^\alpha=-1$.
This formula still contains $\dot{p}^\mu$ and $\dot{S}^{\mu\nu}$, thus naturally depends on solution of the MPD system (which in general cannot be given without adding a ``spin supplementary condition" -- see the following subsection), so it is also {\em not} an explicit momentum-velocity relation, similarly as equations (\ref{p,u,dSu}), (\ref{u,p,dSp}). However, it at least shows clearly that such a closed relation does follow when $S^{\alpha\beta}p_\beta=0$ (Tulczyjew's condition, see section \ref{MP-condition}).

\subsection{Spin supplementary condition}

The effective non-zero size of the ``particle", required by its non-zero multipole moments, implies freedom of its internal motion. On the pole-dipole level, this freedom has 3 degrees and corresponds to a possibility of selecting the representative world-line. A usual choice is to identify the latter with the particle's centre of mass defined with respect to some physical observer. If such an observer is represented by a future-pointing time-like field $V^\mu$, defined ``within the body" (all along its history) and normalized without loss of generality as $V_\sigma V^\sigma=-1$, this means prescribing that the corresponding relative mass dipole is zero, $S^{\mu\sigma}V_\sigma=0$, along that world-line (which is yet to be found, however!). These 3 conditions close the MPD system, the freedom thus being translated into the choice of the reference observer $V^\mu$.

Several specific choices of $V^\mu$ are natural and have proven advantageous, namely $V^\mu\equiv u^\mu$ (Mathisson--Pirani spin condition), $V^\mu\equiv{\cal U}^\mu$ (Tulczyjew's spin condition), $V^\mu=(V^t,0,0,0)$ in a coordinate system adapted to given space-time symmetries (Corinaldesi--Papapetrou spin condition), $V^\mu\equiv{\cal U}^\mu+N^\mu$, where $N^\mu$ is a normalized time-like direction used for 3+1 splitting (Newton--Wigner spin condition, being employed in the Hamiltonian treatment \cite{BarausseRB-09,KunstLLS-15}), and $V^\mu$ given by any parallel vector function along $u^\mu$ (which implies $u^\mu\parallel p^\mu$ and thus $m={\cal M}$, see \cite{KyrianS-07}). Different spin conditions have slightly different character and (naturally) lead to different representative world-lines, which has been stimulating discussions on how uniquely they determine the evolution and whether they actually describe ``the same body"; see \cite{CostaN-15} for a recent thorough treatise on the nature and implications of the different choices.

\subsection{Spin bivector and spin vector}
\label{spin-bivector}

If it satisfies $S^{\mu\sigma}V_\sigma=0$ (its ``electric part" vanishes), the spin bivector must be of rank 2 (must be {\em simple}), having just 3 independent components. With the reference observer $V^\mu$ selected, it is thus possible to introduce a spin vector (``magnetic part" of the bivector) by
\begin{eqnarray}
  &s^\mu& := -\frac{1}{2}\,
              \epsilon^{\mu\nu\rho\sigma}V_\nu S_{\rho\sigma}
           = -{^*\!}S^{\mu\nu}V_\nu  \label{spin-vector} \\
  &\Leftrightarrow&
       \quad S_{\alpha\beta}=\epsilon_{\alpha\beta\gamma\delta}V^\gamma s^\delta,
       \quad {^*\!}S^{\mu\nu}=s^\mu V^\nu-V^\mu s^\nu.
  \label{S-s}
\end{eqnarray}
The vector $s^\mu$ is orthogonal to $V^\mu$ as well as to $S^{\mu\nu}$ by definition, from where also
\begin{equation}  \label{dualS*S=0}
  {^*\!}S^{\mu\iota}S_{\iota\beta}=0 \,.
\end{equation}
In other words, the spin bivector $S^{\mu\nu}$ has two different (orthogonal) eigen-vectors $V^\mu$ and $s^\mu$ tied to zero eigen-values; these vectors span the blade of the dual bivector ${^*\!}S^{\mu\nu}$. Since $V^\mu$ is time-like by assumption, this dual blade is time-like (so ${^*\!}S^{\mu\nu}$ is time-like), hence the blade of $S^{\mu\nu}$, being orthogonal to the dual one, is space-like.

It is useful to also calculate
\begin{align}
  S_{\alpha\beta}S^{\mu\beta}
    &= s^2\left(\delta^\mu_\alpha+V^\mu V_\alpha-s^{-2}s^\mu s_\alpha\right), \\
  -{^*\!}S_{\alpha\beta}{^*\!}S^{\mu\beta}
    &= s^2\left(-V^\mu V_\alpha+s^{-2}s^\mu s_\alpha\right).
\end{align}
Therefore, $-{^*\!}S_{\alpha\beta}{^*\!}S^{\mu\beta}$ represents the $s^2$-multiple of the dual-blade metric, while $S_{\alpha\beta}S^{\mu\beta}$ represents the $s^2$-multiple of the metric of the blade (i.e. that of the surface orthogonal to both $V^\mu$ and $s^\mu$).

The dual blade (the eigen-plane of $S^{\mu\nu}$) can alternatively be spanned, instead of $V^\mu$ and $s^\mu$, by null vectors
\[k^\mu:=\frac{1}{\sqrt{2}}\left(V^\mu+\frac{s^\mu}{s}\right), \quad
  l^\mu:=\frac{1}{\sqrt{2}}\left(V^\mu-\frac{s^\mu}{s}\right),\]
where
\[s^2:=s_\mu s^\mu=\frac{1}{2}\,S_{\alpha\beta}S^{\alpha\beta}
                  =-\frac{1}{2}\,{^*\!}S_{\kappa\lambda}{^*\!}S^{\kappa\lambda}
  \;\;(>0)\]
stands for the spin magnitude squared.
Clearly the vectors are normalized to $k_\mu l^\mu=-1$.
In terms of the null vectors, the bivectors can be expressed
\begin{equation}  \label{bivectors,skl}
  S_{\alpha\beta}=-s\,\epsilon_{\alpha\beta\gamma\delta}k^\gamma l^\delta, \quad
  {^*\!}S^{\mu\nu}=s\,(k^\mu l^\nu-l^\mu k^\nu).
\end{equation}
This implies $S^{\mu\nu}k_\nu=0$ and $S^{\mu\nu}l_\nu=0$,
while ${^*\!}S^{\mu\nu}k_\nu=-s\,k^\mu$ and ${^*\!}S^{\mu\nu}l_\nu=s\,l^\mu$,
so $k^\mu$ and $l^\mu$ are eigen-vectors of ${^*\!}S^{\mu\nu}$ as well (with eigen-values $\mp s$).

The four-momentum $p^\alpha$ was previously extracted from (\ref{Papa-S}) by multiplication by $u_\beta$ or $p_\beta$, but it can now be also expressed in a different way if multiplying the equation by $V_\beta$,
\begin{equation}  \label{p,u,SdV}
  \gamma\,p^\alpha=\mu\,u^\alpha+S^{\alpha\beta}\dot{V}_\beta \,,
\end{equation}
where
\begin{equation}
  \mu:=-V_\sigma p^\sigma \;(>0)\,, \quad
  \gamma:=-V_\sigma u^\sigma \;(>0)
\end{equation}
are, respectively, the particle mass measured with respect to $V^\mu$ and relative Lorentz factor between $u^\mu$ and $V^\mu$. By multiplying this formula once more by $u_\alpha$ or $p_\alpha$, one obtains relations between masses $m$, ${\cal M}$ and $\mu$ (and corresponding projections of $S^{\alpha\beta}$), multiplication by $V_\alpha$ gives just identity, while multiplication respectively by $s_\alpha$, $\dot{V}_\alpha$ and $\dot{s}_\alpha$ yields important equalities
\begin{eqnarray}
  \gamma\,p^\alpha s_\alpha &=& \mu\,u^\alpha s_\alpha \,, \label{ps=us} \\
  \gamma\,p^\alpha \dot{V}_\alpha &=& \mu\,u^\alpha \dot{V}_\alpha \,, \label{pdotV=udotV} \\
  \gamma\,p^\alpha \dot{s}_\alpha &=& \mu\,u^\alpha \dot{s}_\alpha \,. \label{pdots=udots}
\end{eqnarray}
The last one actually follows due to the first two, because thanks to them the $\dot{s}_\alpha$-product of the last term of (\ref{p,u,SdV}) gives zero, too,
\[\dot{s}_\alpha S^{\alpha\beta}\dot{V}_\beta
  =-s_\alpha\dot{S}^{\alpha\beta}\dot{V}_\beta
  =s_\alpha(u^\alpha p^\beta-p^\alpha u^\beta)\dot{V}_\beta
  =0 \,.\]

The above relations tell that the vector $(\mu u^\mu-\gamma p^\mu)$ is orthogonal to $s^\mu$, $\dot{V}^\mu$ and $\dot{s}^\mu$ (and to $V^\mu$ as well). Due to them it is even possible to find a {\em quadruple} of {\em mutually orthogonal} vectors, (e.g.)
\begin{equation}  \label{4D-basis}
  V^\mu, \;\; s^\mu, \;\; \mu u^\mu-\gamma p^\mu, \;\; (s^2\delta^\mu_\nu-s^\mu s_\nu)\,\dot{V}^\nu,
\end{equation}
which can thus be used as a basis (we will indeed use it in section \ref{intrinsic-tetrad}). Another simple space-time basis -- ``built on" $u^\mu$ instead of $V^\mu$ -- will be added in section \ref{eigen-vectors}.
Several simple orthogonal triples can also be found and useful, like
\[\{V^\mu,\;\dot{V}^\mu,\;\mu u^\mu-\gamma p^\mu\}, \quad
  \{u^\mu,\;\mu\dot{u}^\mu-\gamma\dot{p}^\mu,\;\gamma s^\mu+s^\nu u_\nu V^\mu\} \,.\]
Note that the last of these vectors is orthogonal to $p^\mu$ besides, so it is orthogonal to both $u^\mu$ and $p^\mu$, which means that it is an eigen-vector of $\dot{S}^{\mu\nu}$ (with zero eigen-value).

Equations (\ref{ps=us})--(\ref{pdots=udots}) further imply
\begin{align*}
  p^\alpha s_\alpha=0      \quad &\Leftrightarrow \quad u^\alpha s_\alpha=0      \,, \\
  p^\alpha\dot{V}_\alpha=0 \quad &\Leftrightarrow \quad u^\alpha\dot{V}_\alpha=0 \,, \\
  p^\alpha\dot{s}_\alpha=0 \quad &\Leftrightarrow \quad u^\alpha\dot{s}_\alpha=0 \,,
\end{align*}
independently of the spin condition.
Whenever $s^\mu$ {\em is} orthogonal to $u^\mu$ and $p^\mu$, it means that
\[S^{\alpha\beta}\dot{s}_\beta=-\dot{S}^{\alpha\beta}s_\beta=0 \,,\]
so $\dot{s}^\mu$ then belongs to the eigen-plane of $S^{\mu\nu}$ -- it is a combination of $V^\mu$ and $s^\mu$. Conversely, $s^\mu$ belongs then to the eigen-plane of $\dot{S}^{\mu\nu}$, the other independent eigen-direction of the latter being given by $\epsilon^{\mu\iota\kappa\lambda}s_\iota u_\kappa p_\lambda$.
Similarly, when $\dot{V}^\mu$ is orthogonal to $u^\mu$ and $p^\mu$, it means that $\dot{S}^{\alpha\beta}\dot{V}_\beta=0$, so $\dot{V}^\mu$ is the eigen-vector of $\dot{S}^{\mu\nu}$, the other one being $\epsilon^{\mu\iota\kappa\lambda}\dot{V}_\iota u_\kappa p_\lambda$.
And, finally, when $\dot{s}^\mu$ is orthogonal to $u^\mu$ and $p^\mu$, it means that $\dot{S}^{\alpha\beta}\dot{s}_\beta=0$, so $\dot{s}^\mu$ is the eigen-vector of $\dot{S}^{\mu\nu}$, the other one being $\epsilon^{\mu\iota\kappa\lambda}\dot{s}_\iota u_\kappa p_\lambda$.

The above reasoning is clearly pointless if $u^\mu$ is parallel to $p^\mu$; this circumstance will be more discussed in section \ref{our-condition}.

\subsubsection{Hidden momentum}

Inspired by the concept of ``hidden momentum" used in electromagnetism, \cite{GrallaHW-10} introduced its ``gravitational" counterpart analogously as the component of $p^\mu$ orthogonal to $u^\mu$. Since we assume the particle is torque-free, it is solely given by the chosen spin supplementary condition in our case (it is purely kinematical, see \cite{CostaN-15} for details),
\begin{align}
  p^\mu_{\rm hidden} &:=(\delta^\mu_\alpha+u^\mu u_\alpha)\,p^\alpha
                       =p^\mu-mu^\mu = \\
                     & =-\dot{S}^{\mu\nu}u_\nu
                       =\frac{1}{\gamma}\,(\delta^\mu_\alpha+u^\mu u_\alpha)\,
                        S^{\alpha\beta}\dot{V}_\beta \,.  \label{hidden-p}
\end{align}
We will refer to this term in section \ref{our-condition} where the option of making $p^\mu_{\rm hidden}=0$ will be discussed.

\subsection{MPD equations in terms of spin vector}

Writing out the left-hand side of (\ref{Papa-S}) in terms of $s^\mu$,
\begin{equation}  \label{dotS-dots}
  \dot{S}_{\alpha\beta}=
  \epsilon_{\alpha\beta\gamma\delta}\dot{V}^\gamma s^\delta+
  \epsilon_{\alpha\beta\gamma\delta}V^\gamma\dot{s}^\delta,
\end{equation}
and then multiplying the equation by $\epsilon^{\mu\nu\alpha\beta}V_\nu$, one has
\begin{equation}  \label{dots,project}
  (\delta^\mu_\nu+V^\mu V_\nu)\,\dot{s}^\nu
  =\epsilon^{\mu\nu\alpha\beta}V_\nu u_\alpha p_\beta
\end{equation}
and hence
\begin{equation}  \label{dots}
  \dot{s}^\mu=V^\mu\dot{V}_\nu s^\nu+\epsilon^{\mu\nu\alpha\beta}V_\nu u_\alpha p_\beta \,.
\end{equation}
Therefore, the change of spin along $u^\mu$ is parallel to $V^\mu$ in two obvious cases:
(i) whenever $u^\mu$ is parallel to $p^\mu$ (which implies $\dot{S}_{\alpha\beta}=0$ and ${\cal M}=m$);
(ii) if $V^\mu$ lies in the plane spanned by $u^\mu$ and ${\cal U}^\mu$ (i.e., if one applies some combination of the Mathisson--Pirani and Tulczyjew conditions).
The spin magnitude evolves according to
\begin{eqnarray}
  s\dot{s} \equiv s\,\frac{{\rm d}s}{{\rm d}\tau}
   = \frac{1}{2}\,\frac{{\rm d}s^2}{{\rm d}\tau}
  &=&\frac{1}{2}\,S^{\alpha\beta}\dot{S}_{\alpha\beta}
   = S^{\alpha\beta}p_\alpha u_\beta \\
  &=&s_\mu\dot{s}^\mu
   = \epsilon^{\mu\nu\alpha\beta}s_\mu V_\nu u_\alpha p_\beta \,, \label{sds=EsVup}
\end{eqnarray}
which in the above two cases yields conservation.
Note in passing that $s_\mu(s\dot{s}^\mu-\dot{s}s^\mu)=0$, so, regarding (\ref{ps=us}) and (\ref{pdots=udots}), the vectors
\[s^\mu, \quad s\dot{s}^\mu-\dot{s}s^\mu, \quad \mu u^\mu-\gamma p^\mu\]
are orthogonal to each other.

Similarly, by multiplying the relation (\ref{dotS-dots}) by $\epsilon^{\mu\nu\alpha\beta}s_\nu$ and using (\ref{sds=EsVup}), one arrives at
\begin{equation}  \label{dotV,project}
  (s^2\delta^\mu_\nu-s^\mu s_\nu)\,\dot{V}^\nu
  =(\delta^\mu_\nu+V^\mu V_\nu)\,\epsilon^{\nu\iota\alpha\beta}s_\iota u_\alpha p_\beta
\end{equation}
and hence
\begin{equation}  \label{dotV}
  s\,\frac{{\rm D}}{{\rm d}\tau}(s V^\mu)=
  -s^\mu\dot{s}_\nu V^\nu
  +\epsilon^{\mu\iota\alpha\beta}s_\iota u_\alpha p_\beta \,.
\end{equation}

Introducing (\ref{dotS-dots}) and then (\ref{dots}) into the mass-evolution formulas (\ref{dm})--(\ref{dM}), we obtain, after some rearrangement,
\begin{eqnarray}
  \gamma^2\dot{m}&=&
  \epsilon^{\alpha\beta\rho\sigma}s_\alpha\dot{V}_\beta u_\rho
  \dot{u}_\iota(\delta^\iota_\sigma+V^\iota V_\sigma) \,, \label{dotm} \\
  \gamma^2{\cal M}\dot{\cal M}&=&
  \epsilon^{\alpha\beta\rho\sigma}s_\alpha\dot{V}_\beta u_\rho
  \dot{p}_\iota(\delta^\iota_\sigma+V^\iota V_\sigma) \,. \label{dotM}
\end{eqnarray}
In particular, the $V^\mu\equiv u^\mu$ choice leads to $\dot{m}=0$, while the $V^\mu\equiv{\cal U}^\mu$ choice leads to $\dot{\cal M}=0$.
On the other hand, the evolution of $\mu\equiv -V_\sigma p^\sigma$ can be expressed, using (\ref{pdotV=udotV}), as
\begin{equation}  \label{dotmu}
  \gamma\,\dot{\mu}
  = -\gamma\,\dot{V}_\lambda p^\lambda-\gamma\,V_\lambda\dot{p}^\lambda
  = -\mu\,u_\kappa\dot{V}^\kappa-\gamma\,V_\lambda\dot{p}^\lambda \,.
\end{equation}
Naturally, this reduces to the above limits too.

Now to the main point at last: written in terms of the spin vector, the first MPD equation appears as
\begin{eqnarray}
  \dot{p}^\mu
  &=& -\frac{1}{2}\,{R^\mu}_{\nu\kappa\lambda}u^\nu S^{\kappa\lambda}
   =  -\frac{1}{2}\,g^{\mu\rho}R_{\rho\nu\kappa\lambda}
       \epsilon^{\kappa\lambda\alpha\beta}u^\nu V_\alpha s_\beta \nonumber \\
  &=& -g^{\mu\rho}R^*_{\rho\nu\alpha\beta}u^\nu V^\alpha s^\beta
   =  {^*\!R^\mu}_{\nu\alpha\beta}u^\nu s^\alpha V^\beta,  \label{Dp,s}
\end{eqnarray}
where $R^*_{\rho\nu\alpha\beta}$ and ${^*\!R}_{\rho\nu\alpha\beta}$ are the Riemann-tensor right and left duals; in the last equality, we have used the fact that they are equal in the vacuum case (even with non-zero cosmological constant).

\subsection{Eigen-vectors of $S^{\mu\nu}$, ${^*\!}S^{\mu\nu}$, $\dot{S}^{\mu\nu}$, ${^*\!}\dot{S}^{\mu\nu}$}
\label{eigen-vectors}

Already at several places we have mentioned eigen-vectors of the spin bivector and of its derivative and/or dual. All the bivectors indeed have the whole 2D eigen-planes (corresponding to zero eigen-values) which can be spanned in a number of ways, so let us just summarize one simple possibility for each.\\
(i) $S_{\alpha\beta}=\epsilon_{\alpha\beta\mu\nu}V^\mu s^\nu$,
so it is clear that $V^\beta$ and $s^\beta$ are its ``default" eigen-vectors -- they are simple and orthogonal to each other.\\
(ii) ${^*\!}S^{\mu\nu}=s^\mu V^\nu-V^\mu s^\nu$.
Thanks to the property (\ref{dualS*S=0}), the eigen-vectors of ${^*\!}S^{\mu\nu}$ can be found simply by any non-trivial projections of $S_{\nu\beta}$. One suitable vector for such a projection is $\dot{V}^\beta$,
\[-S_{\nu\beta}\dot{V}^\beta=\dot{S}_{\nu\beta}V^\beta
  =(p_\nu u_\beta-u_\nu p_\beta)\,V^\beta
  =\mu u_\nu-\gamma p_\nu \,.\]
We would like the second eigen-vector to be orthogonal to the one just found, which is for example true for the vector
(\ref{dotV,project}). The latter is in fact orthogonal to all the vectors $V^\mu$, $s^\mu$, $p^\mu$ and $u^\mu$, so it clearly has both the desired properties.
Note that actually {\em all} the eigen-vectors suggested in (i) and (ii) are mutually orthogonal: sure, they just form the basis we already know from (\ref{4D-basis}).\\
(iii) ${^*\!}\dot{S}_{\mu\nu}=\epsilon_{\mu\nu\alpha\beta}p^\alpha u^\beta$,
so the ``default" eigen-vectors seem to be $p^\nu$ and $u^\nu$. These two are however {\em never} orthogonal, yet it is easy to fix this by taking
\[u^\nu \qquad {\rm and} \qquad
  (\delta^\nu_\sigma+u^\nu u_\sigma)\,p^\sigma=p^\nu-mu^\nu \; (=-\dot{S}^{\nu\beta}u_\beta).\]
The last vector is exactly the ``hidden momentum" which turned out to be useful in understanding of the pole-dipole description \cite{CostaN-15}. \\
(iv) $\dot{S}^{\alpha\beta}=p^\alpha u^\beta-u^\alpha p^\beta$.
One finds shortly that one eigen-vector, moreover orthogonal to both $u^\beta$ and $p^\beta$ (hence also to $p^\beta\!-\!mu^\beta$), is $(\gamma s_\beta+s_\nu u^\nu V_\beta)$. The second one, orthogonal to the latter as well as to $u^\beta$ and $p^\beta-mu^\beta$, reads
\begin{align}
  &\epsilon^{\beta\iota\kappa\lambda}u_\iota p_\kappa(\gamma s_\lambda+s_\nu u^\nu V_\lambda) = \nonumber \\
  &= \gamma s\dot{s}V^\beta
     +\gamma(s^2\delta^\beta_\nu\!-\!s^\beta s_\nu)\,\dot{V}^\nu
     +u^\sigma s_\sigma(\delta^\beta_\nu\!+\!V^\beta V_\nu)\,\dot{s}^\nu.
\end{align}

Clearly the eigen-vector choices suggested for the (ii)--(iv) bivectors are only realizable in generic case, in particular, they are not valid if $p^\mu=mu^\mu$ when most of them become trivial. This special ``gauge" will be discussed below, along with the other two major ones (section \ref{SSCs}).

\section{Vacuum MPD equations in a suitable tetrad}
\label{MPD-in-tetrad}

In a vacuum region, yet possibly involving a non-zero cosmological constant $\Lambda$, the usual decomposition of the Riemann tensor reduces to
\begin{equation}
  R_{\mu\nu\kappa\lambda}=
  C_{\mu\nu\kappa\lambda}
  +\frac{\Lambda}{3}\,(g_{\mu\kappa}g_{\nu\lambda}-g_{\mu\lambda}g_{\nu\kappa}) \,,
\end{equation}
where $C_{\mu\nu\kappa\lambda}$ denotes the Weyl tensor.
In some orthonormal tetrad $\{e_{\hat\alpha}^\mu\}$ ($\alpha$ numbers the vectors), $g_{\mu\nu}e_{\hat\alpha}^\mu e_{\hat\beta}^\nu=\eta_{\alpha\beta}$, the decomposition reads
\begin{equation}
  R_{\hat\alpha\hat\beta\hat\gamma\hat\delta}=
  C_{\hat\alpha\hat\beta\hat\gamma\hat\delta}
  +\frac{\Lambda}{3}\,(\eta_{\alpha\gamma}\eta_{\beta\delta}-\eta_{\alpha\delta}\eta_{\beta\gamma}) \,,
\end{equation}
where $\eta_{\alpha\beta}$ is the Minkowski tensor.
The dual tensor thus decomposes as
\begin{equation}  \label{*R-decomp}
  {^*\!R^\mu}_{\nu\alpha\beta}
  ={^*\!C^\mu}_{\nu\alpha\beta}+\frac{\Lambda}{3}\,{\epsilon^\mu}_{\nu\alpha\beta} \,.
\end{equation}

\subsection{Orthonormal tetrad}
\label{ortho-basis}

When considering the choice of a suitable tetrad, one has 3 logical options -- $u^\mu$, ${\cal U}^\mu$ and $V^\mu$ -- for the time-like vector $e^\mu_{\hat{0}}$. The choice $e^\mu_{\hat{0}}\equiv V^\mu$ and $e^\mu_{\hat{1}}\equiv s^\mu/s$ seems the most advantageous and ``universal", because
\begin{itemize}
\item
the spin $s^\mu$, fixed by $V^\mu$ and introduced as orthogonal to the latter in every case, represents the most important space-like direction of the problem, so it is natural to discuss the curvature effects with respect to it
\item
the reference observer $V^\mu$ is actually the only vector that one chooses freely, the other ones depending on it: $V^\mu$ fixes the meaning of $p^\mu$ and $S^{\mu\nu}$ as moments of the energy-momentum tensor by specifying the hypersurface over which they are calculated,\footnote
{However, it is worth to note that in the pole-dipole order and with just gravitational effects included, the $p^\mu$ is in fact same for {\em any} spin condition.}
as well as that of $u^\mu$ identified as the tangent to the world-line along which $S^{\mu\sigma}V_\sigma=0$ holds
\item
the freedom in $V^\mu$ makes it easily adaptable to particular space-times and situations, specifically to the particular Petrov types
\item
special pictures arising for particular spin supplementary conditions follow simply by selecting $V^\mu$ accordingly, e.g. as $u^\mu$, ${\cal U}^\mu$ or as some vector parallel along $u^\mu$.
\end{itemize}
Let us note that it might seem preferable to choose $u^\mu$ as {\em the} time-like direction, for it is (its finding is) certainly central to the problem, and because $u_\mu\dot{p}^\mu=0$ which makes the time component of the problem settled immediately. However, such a choice does not allow to select the spin $s^\mu$ as one of the basis directions, because $s^\mu$ is orthogonal to $V^\mu$ rather than to $u^\mu$ (the latter would require $u_\mu s^\mu=u_\mu V_\nu {^*\!}S^{\nu\mu}=0$ which is not the case in general). And, obviously, $u^\mu$ cannot be {\em chosen}, it is given by the MPD equations and a chosen spin condition.

Having opted for $e^\mu_{\hat{0}}\equiv V^\mu$, $e^\mu_{\hat{1}}\equiv s^\mu/s$, we have from (\ref{Dp,s}) and (\ref{*R-decomp})
\begin{eqnarray}
  \dot{p}^\mu &=& {^*\!R^\mu}_{\nu\alpha\beta}u^\nu s^\alpha V^\beta
               =  s\,{^*\!R^\mu}_{\hat{\gamma}\hat{1}\hat{0}}u^{\hat{\gamma}}= \nonumber \\
              &=& s\,{^*\!C^\mu}_{\hat{\gamma}\hat{1}\hat{0}}u^{\hat{\gamma}}
                  +\frac{\Lambda}{3}\,s\,{\epsilon^\mu}_{\hat{\gamma}\hat{1}\hat{0}}u^{\hat{\gamma}} \,.
  \label{Dp,s,hat}
\end{eqnarray}
The projections of equation (\ref{Dp,s,hat}) on $V^\mu$ and $e^\mu_{\hat{1}}$ read, respectively,
\begin{eqnarray}
  -V_\mu\,\dot{p}^\mu
    &=& \dot{\mu}+p_\mu\dot{V}^\mu=\dot{\mu}+p_{\hat{k}}\dot{V}^{\hat{k}}
     =  s\,{^*\!C}_{\hat{\imath}\hat{0}\hat{1}\hat{0}}u^{\hat{\imath}}, \\
  e^{\hat{1}}_\mu\,\dot{p}^\mu
    &=& s\,{^*\!C}_{\hat{1}\hat{\gamma}\hat{1}\hat{0}}u^{\hat{\gamma}},
\end{eqnarray}
where the notation $\mu:=-V_\sigma p^\sigma$ has been recalled. The cosmological term has no effect in this plane.
The remaining two spatial directions $e^\mu_{\hat{2}}$ and $e^\mu_{\hat{3}}$, perpendicular to both $V^\mu$ and $s^\mu$, are left unspecified for the moment; the respective projections of (\ref{Dp,s,hat}) write
\begin{eqnarray}
  e^{\hat{2}}_\mu\,\dot{p}^\mu
    &=& s\,{^*\!C}_{\hat{2}\hat{\gamma}\hat{1}\hat{0}}u^{\hat{\gamma}}
        -\frac{\Lambda}{3}\,s\,u^{\hat{3}} \,, \\
  e^{\hat{3}}_\mu\,\dot{p}^\mu
    &=& s\,{^*\!C}_{\hat{3}\hat{\gamma}\hat{1}\hat{0}}u^{\hat{\gamma}}
        +\frac{\Lambda}{3}\,s\,u^{\hat{2}} \,.
\end{eqnarray}
Hence, the time component $-V_\mu\,\dot{p}^\mu$ is determined purely by the magnetic part of the Weyl tensor, $B_{\hat{\imath}\hat{1}}u^{\hat{\imath}}:={^*\!C}_{\hat{\imath}\hat{0}\hat{1}\hat{0}}u^{\hat{\imath}}$ (see appendix \ref{Weyl-electric,magnetic}), while the remaining components are influenced by both magnetic terms (those containing $u^{\hat{0}}:=-V_\sigma u^\sigma\equiv\gamma$) and electric terms (containing $u^{\hat{\imath}}$).

\subsection{Newman--Penrose null tetrad}

Let us proceed now to the standard Newman--Penrose tetrad made of two real and two complex null vectors $(k^\mu,l^\mu,m^\mu,\bar{m}^\mu)$ introduced by
\begin{eqnarray}
  k^\mu:=\frac{1}{\sqrt{2}}\,(V^\mu+e^\mu_{\hat{1}}), &\quad&
  l^\mu:=\frac{1}{\sqrt{2}}\,(V^\mu-e^\mu_{\hat{1}}), \\
  m^\mu:=\frac{1}{\sqrt{2}}\,(e^\mu_{\hat{2}}+{\rm i}\,e^\mu_{\hat{3}}), &\quad&
  \bar{m}^\mu:=\frac{1}{\sqrt{2}}\,(e^\mu_{\hat{2}}-{\rm i}\,e^\mu_{\hat{3}}),
\end{eqnarray}
satisfying the normalizations
\begin{eqnarray*}
  k_\mu l^\mu=-1, &\quad&
  m_\mu \bar{m}^\mu=1, \\
  k_\mu m^\mu=k_\mu\bar{m}^\mu &=&
  l_\mu m^\mu=l_\mu\bar{m}^\mu=0
\end{eqnarray*}
and giving rise to the metric decomposition
\begin{equation}
  g_{\mu\nu}=-k_\mu l_\nu-k_\nu l_\mu+m_\mu\bar{m}_\nu+m_\nu\bar{m}_\mu \,.
\end{equation}
The 10 independent components of the Weyl tensor are represented by 5 independent complex projections
\begin{eqnarray}
  \Psi_0 &:=& C_{\mu\nu\kappa\lambda}k^\mu m^\nu k^\kappa m^\lambda, \\
  \Psi_1 &:=& C_{\mu\nu\kappa\lambda}k^\mu l^\nu k^\kappa m^\lambda, \\
  \Psi_2 &:=& C_{\mu\nu\kappa\lambda}k^\mu m^\nu\bar{m}^\kappa l^\lambda \\
         & =&  \frac{1}{2}\,C_{\mu\nu\kappa\lambda}
              (k^\mu l^\nu k^\kappa l^\lambda-k^\mu l^\nu m^\kappa\bar{m}^\lambda), \\
  \Psi_3 &:=& C_{\mu\nu\kappa\lambda}l^\mu k^\nu l^\kappa\bar{m}^\lambda, \\
  \Psi_4 &:=& C_{\mu\nu\kappa\lambda}l^\mu\bar{m}^\nu l^\kappa\bar{m}^\lambda.
\end{eqnarray}

We need to find how the above scalars relate to their counterparts given by the {\em dual} Weyl tensor now. It is clear from the latter's definition that the dualization can be shifted to the respective couple of tetrad vectors in the projections, so it is sufficient to find how the bivectors made of the tetrad elements behave under dualization; actually, due to the symmetries of the (dual) Weyl tensor, it is sufficient to know this for $k^\mu m^\nu$ and $l^\mu\bar{m}^\nu$. It is easy to check that the null tetrad is ``positively" oriented,
\[\epsilon_{\alpha\beta\gamma\delta}k^\alpha l^\beta m^\gamma \bar{m}^\delta
  ={\rm i}\,[0123]={\rm i} \,,\]
and that the Hodge star simply brings imaginary unit,
\begin{equation}
  {^*}(k^\mu\wedge m^\nu)={\rm i}\,(k^\mu\wedge m^\nu), \quad
  {^*}(l^\mu\wedge\bar{m}^\nu)={\rm i}\,(l^\mu\wedge\bar{m}^\nu),
\end{equation}
which implies ``anti-self-duality" of the scalars,
\begin{eqnarray}
  {^*}\Psi_0 &:=& {^*\!C}_{\mu\nu\kappa\lambda}k^\mu m^\nu k^\kappa m^\lambda
         = {\rm i}\,\Psi_0 \,, \\
  {^*}\Psi_1 &:=& {^*\!C}_{\mu\nu\kappa\lambda}k^\mu l^\nu k^\kappa m^\lambda
         = {\rm i}\,\Psi_1 \,, \\
  {^*}\Psi_2 &:=& {^*\!C}_{\mu\nu\kappa\lambda}k^\mu m^\nu\bar{m}^\kappa l^\lambda
         = {\rm i}\,\Psi_2 \,, \\
  {^*}\Psi_3 &:=& {^*\!C}_{\mu\nu\kappa\lambda}l^\mu k^\nu l^\kappa\bar{m}^\lambda
         = {\rm i}\,\Psi_3 \,, \\
  {^*}\Psi_4 &:=& {^*\!C}_{\mu\nu\kappa\lambda}l^\mu\bar{m}^\nu l^\kappa\bar{m}^\lambda
         = {\rm i}\,\Psi_4 \,.
\end{eqnarray}
Writing, conversely,
\begin{eqnarray}
  V^\mu=\frac{1}{\sqrt{2}}\,(k^\mu+l^\mu), &\quad&
  e^\mu_{\hat{1}}=\frac{1}{\sqrt{2}}\,(k^\mu-l^\mu), \label{V,e1} \\
  e^\mu_{\hat{2}}=\frac{1}{\sqrt{2}}\,(m^\mu+\bar{m}^\mu), &\quad&
  e^\mu_{\hat{3}}=\frac{1}{\sqrt{2}\,{\rm i}}\,(m^\mu-\bar{m}^\mu), \label{e2,e3}
\end{eqnarray}
and using Appendix \ref{Appendix-A}, we thus obtain
\begin{eqnarray}
  -V_\mu\,\dot{p}^\mu
     &=& -2s\,{\rm Im}\Psi_2\,u^{\hat{1}} \nonumber \\
     &{}& -s\,({\rm Im}\Psi_3-{\rm Im}\Psi_1)\,u^{\hat{2}} \nonumber \\
     &{}& -s\,({\rm Re}\Psi_3+{\rm Re}\Psi_1)\,u^{\hat{3}} \,,
     \label{V,dotp} \\
  e^{\hat{1}}_\mu\,\dot{p}^\mu
     &=& -2s\,{\rm Im}\Psi_2\,u^{\hat{0}} \nonumber \\
     &{}& -s\,({\rm Im}\Psi_3+{\rm Im}\Psi_1)\,u^{\hat{2}} \nonumber \\
     &{}& -s\,({\rm Re}\Psi_3-{\rm Re}\Psi_1)\,u^{\hat{3}} \,,
     \label{e1,dotp} \\
  e^{\hat{2}}_\mu\,\dot{p}^\mu
     &=& +s\left(2\,{\rm Re}\Psi_2-\frac{\Lambda}{3}\right)u^{\hat{3}} \nonumber \\
     &{}& -s\,({\rm Im}\Psi_3-{\rm Im}\Psi_1)\,u^{\hat{0}} \nonumber \\
     &{}& +s\,({\rm Im}\Psi_3+{\rm Im}\Psi_1)\,u^{\hat{1}} \,, \\
  e^{\hat{3}}_\mu\,\dot{p}^\mu
     &=& -s\left(2\,{\rm Re}\Psi_2-\frac{\Lambda}{3}\right)u^{\hat{2}} \nonumber \\
     &{}& -s\,({\rm Re}\Psi_3+{\rm Re}\Psi_1)\,u^{\hat{0}} \nonumber \\
     &{}& +s\,({\rm Re}\Psi_3-{\rm Re}\Psi_1)\,u^{\hat{1}} \,.
     \label{e3,dotp}
\end{eqnarray}
The main feature of these equations is that they do not at all contain $\Psi_0$ and $\Psi_4$. And note again that the cosmological constant only influences motion in the spatial directions perpendicular to spin.

It may seem possible to express the above equations in terms of only two components of four-velocity, because $u^{\hat{\alpha}}$ are constrained by the relation
\[0=u_\mu\dot{p}^\mu
   =u^{\hat{0}}V_\mu\dot{p}^\mu+\delta_{ij}u^{\hat{\imath}}e^{\hat{\jmath}}_\mu\dot{p}^\mu\]
and by normalization
\[-1=-(u^{\hat{0}})^2+\delta_{ij}u^{\hat{\imath}}u^{\hat{\jmath}}.\]
But $u_\mu\dot{p}^\mu=0$ is satisfied {\em automatically}, it brings no information (it can actually be used as a correctness check).

\subsection{Simple implications from tetrad components of the MPD equation of motion}

It should first be stressed that $u_\mu\dot{p}^\mu=0$, so the overall effect is {\em always} perpendicular to the representative world-line, irrespective of any interpretation superstructure.
In the above projections, the simplest terms are the ``cosmological" ones: they act within the $(e_{\hat{2}}^\mu,e_{\hat{3}}^\mu)$ plane and always perpendicular to the projection of the trajectory onto this plane. The ``Newton--Coulomb" tidal field, generated by mass and given by ${\rm Re}\Psi_2$, acts in the same manner within the same plane, its effect having the opposite/same orientation for positive/negative $\Lambda$. The ``magnetic-type" tidal field, generated by angular momentum and given by ${\rm Im}\Psi_2$, acts in exactly the same way, but within the orthogonal $(e_{\hat{0}}^\mu,e_{\hat{1}}^\mu)$ plane. The remaining force is tied to longitudinal wave effects in the $k^\mu$ and $l^\mu$ directions, represented by $\Psi_1$ and $\Psi_3$ (cf. section \ref{transverse}). The last scalars $\Psi_0$ and $\Psi_4$ are not at all present, which in standard understanding (cf. \cite{HofmannNS-13}) means that if the NP tetrad is chosen as we did, there are no transverse wave effects, neither in $k^\mu$ nor in $l^\mu$ direction.

\subsection{``Intrinsic" choice of tetrad}
\label{intrinsic-tetrad}

It was notably Ernst Mach who emphasized that the system should itself provide terms in which it will be interpreted. Unfortunately, our system needs the ``reference observer" $V^\mu$ in order to be unique and to make sense. However, we know from (\ref{4D-basis}) that with this chosen, the spinning particle does provide a unique orthogonal (thus also orthonormal) basis which can be used in generic situation (namely when the vectors $p^\mu$ and $u^\mu$ are independent): besides $V^\mu$ and $s^\mu$ from which we have started, it is given by $(\mu u^\mu-\gamma p^\mu)$ and by the ``vector product" of these three,
\begin{align}
  &\epsilon^{\mu\iota\kappa\lambda}V_\iota s_\kappa(\mu u_\lambda-\gamma p_\lambda)
   =S^{\mu\lambda}(\mu u_\lambda-\gamma p_\lambda)= \nonumber \\
  &=-S^{\mu\lambda}S_{\lambda\nu}\dot{V}^\nu
   =(s^2\delta^\mu_\nu-s^\mu s_\nu)\,\dot{V}^\nu\,;
\end{align}
note again that we already mentioned this basis in (\ref{4D-basis}).
Hence, besides the $-V_\mu u^\mu\equiv\gamma\equiv u^{\hat{0}}$ and $s_\mu u^\mu\equiv su^{\hat{1}}$ four-velocity components, we have
\begin{align}
  (\mu u_\mu-\gamma p_\mu)\,u^\mu &= \gamma m-\mu \,, \\
  \epsilon_{\mu\iota\kappa\lambda}V^\iota s^\kappa(\mu u^\lambda-\gamma p^\lambda)\,u^\mu
   &= -\gamma\,\epsilon_{\mu\iota\kappa\lambda}u^\mu V^\iota s^\kappa p^\lambda= \nonumber \\
   &= \gamma\,s\dot{s} \,.
\end{align}
The 2nd and the 3rd components of $\dot{p}^\mu$ follow from
\begin{align}
  (\mu u_\mu-\gamma p_\mu)\,\dot{p}^\mu
  &= -\gamma p_\mu\dot{p}^\mu = \gamma{\cal M}\dot{\cal M} \,, \\
  \epsilon_{\mu\iota\kappa\lambda}V^\iota s^\kappa(\mu u^\lambda-\gamma p^\lambda)\,\dot{p}^\mu
  &= -\dot{p}^\mu S_{\mu\lambda}S^{\lambda\nu}\dot{V}_\nu= \nonumber \\
  &= (s^2 g_{\mu\nu}-s_\mu s_\nu)\,\dot{p}^\mu\dot{V}^\nu.
\end{align}
Before usage of the above tetrad, one should learn the norm of the two newly specified vectors,
\begin{align}
  &(\mu u_\mu\!-\gamma p_\mu)(\mu u^\mu\!-\gamma p^\mu)
   = 2\gamma\mu m-\mu^2-\gamma^2{\cal M}^2 \,, \\
  &\epsilon_{\mu\alpha\beta\gamma}V^\alpha s^\beta(\mu u^\gamma\!-\gamma p^\gamma)\;
   \epsilon^{\mu\iota\kappa\lambda}V_\iota s_\kappa(\mu u_\lambda\!-\gamma p_\lambda)=
   \nonumber \\
  &\qquad\qquad\qquad\qquad =s^2\,(2\gamma\mu m-\mu^2-\gamma^2{\cal M}^2) \,.
\end{align}
Note that the expression crucial in both these relations becomes
\[2\gamma\mu m-\mu^2-\gamma^2{\cal M}^2
  \;\; \longrightarrow\;\;
  m^2-{\cal M}^2 \;(\,\geq 0)\]
for the Mathisson--Pirani as well as the Tulczyjew condition.

Just to summarize, in an {\em orthonormal} tetrad involving the above vectors, one has
\begin{align}
  u^{\hat{2}}:=e^{\hat{2}}_\mu\,u^\mu
              &= \frac{\gamma\,m-\mu}
                      {\sqrt{2\gamma\mu m-\mu^2-\gamma^2{\cal M}^2}} \;,
              \label{u2,intrinsic} \\
  u^{\hat{3}}:=e^{\hat{3}}_\mu\,u^\mu
              &= \frac{\gamma\,\dot{s}}
                      {\sqrt{2\gamma\mu m-\mu^2-\gamma^2{\cal M}^2}} \;, \\
  e^{\hat{2}}_\mu\,\dot{p}^\mu
              &= \frac{\gamma\,{\cal M}\dot{\cal M}}
                      {\sqrt{2\gamma\mu m-\mu^2-\gamma^2{\cal M}^2}} \;, \\
  e^{\hat{3}}_\mu\,\dot{p}^\mu
              &= \frac{(s^2 g_{\mu\nu}-s_\mu s_\nu)\,\dot{p}^\mu\dot{V}^\nu}
                      {s\,\sqrt{2\gamma\mu m-\mu^2-\gamma^2{\cal M}^2}} \;.
              \label{dotp3,intrinsic}
\end{align}

Let us stress that the above tetrad is quite generic (it only cannot be used with the spin condition $u^\mu\parallel p^\mu$ which implies $\mu u^\mu\!-\gamma p^\mu=0$), in particular, it is independent of the space-time curvature structure.

\section{Equation of motion in space-times of different algebraic types}
\label{MPD,Petrov-types}

Let us look whether the above projections of the MPD equation of motion can be somehow linked to the algebraic type of curvature. This mainly means to turn the interpretation tetrad so as to reflect both the spin and the curvature features. First we leave the supplementary spin condition generic, then we will consider the mostly used special choices.

\subsection{Interpretation in a tetrad tied to curvature structure}
\label{tetrad-construction}

One of the aims of this paper is to check whether in some cases the above structure tied to the particle spin can ``fit in" space-time curvature in such a manner that the MPD equations assume an especially simple and easily interpretable form. It is known that every space-time hosts four null eigen-directions of the Weyl tensor, called its principal null directions (PNDs); their multiplicities indicate the Petrov type of space-time. Let us denote by $k^\mu$ the PND of the highest multiplicity. Let us also define another independent real null vector $l^\mu$ and normalize it by $k_\mu l^\mu=-1$; it can in general be chosen arbitrarily (it even {\em has to be} if the curvature is of type N and $k^\mu$ is a four-fold PND), though it would be especially beneficial to take as $l^\mu$ some of the other PNDs. However, such a choice is typically not feasible, if one needs to also simultaneously fix the tetrad to the spin structure.

The only way how to ``adapt" the spin structure to some space-time features is to choose the reference observer $V^\mu$ accordingly and thus adjust the spin eigen-space in some preferred direction. Consider what can be achieved in this respect at some point of the particle world-line. Let us have some space-time and use its highest-multiplicity PND $k^\mu$ as the first vector of the NP tetrad. Now, imagine first that one knows the particle's spin vector $s^\mu$. Then it is natural to define the second null vector of the NP tetrad by
\[l^\mu := \frac{s^2\,k^\mu}{2(s_\iota k^\iota)^2} - \frac{s^\mu}{s_\iota k^\iota} \;.\]
It is easy to check that such $l^\mu$ is really null and that $k_\mu l^\mu=-1$.
Clearly the definition is not possible if $s_\iota k^\iota=0$. Supposing that this is not the case and regarding that a null vector can be normalized arbitrarily, let us choose the normalization of $k^\mu$ so that $s_\iota k^\iota=\frac{s}{\sqrt{2}}$. Then we can specify the definition of $l^\mu$ to
\begin{equation}  \label{k,s->l}
  l^\mu := k^\mu-\sqrt{2}\,\frac{s^\mu}{s} \,,
\end{equation}
which inverts to $s^\mu=\frac{s}{\sqrt{2}}\,(k^\mu-l^\mu)$.
Subsequently, choose the reference observer by
\[V^\mu:= \frac{1}{\sqrt{2}}\,(k^\mu+l^\mu)\]
and introduce the spin bivector by
\[S_{\alpha\beta}=\epsilon_{\alpha\beta\gamma\delta}V^\gamma s^\delta
                 =-s\,\epsilon_{\alpha\beta\gamma\delta}k^\gamma l^\delta.\]

As the second situation, imagine -- as is the case when starting from the original MPD equations -- that one knows the particle's spin bivector $S^{\mu\nu}$ (rather than the vector $s^\mu$). Then it is natural to define the second null vector $l^\mu$ of the NP tetrad as {\em some} null eigen-vector of $S^{\mu\nu}$, i.e. as a vector satisfying $S_{\mu\nu}l^\nu=0$, and normalize it by $k_\mu l^\mu=-1$. (If $l^\mu$ happened to be proportional to $k^\mu$, one would have to take the other independent null eigen-vector of $S^{\mu\nu}$.) Having both $k^\mu$ and $l^\mu$, one chooses
\[V^\mu:= \frac{1}{\sqrt{2}}\,(k^\mu+l^\mu), \quad
  s^\mu=\frac{s}{\sqrt{2}}\,(k^\mu-l^\mu)\]
as the reference observer and the corresponding spin vector.
Since $S_{\mu\nu}V^\nu=0$ according to the required spin condition, the eigen-plane of $S^{\mu\nu}$ is thus turned so as to also contain $k^\mu$, with $s^\mu$ lying in it automatically.

Finally, the plane orthogonal to both $k^\mu$ and $l^\mu$ can  be spanned by the remaining two complex null vectors $m^\mu$ and $\bar{m}^\mu$ arbitrarily (turning these suitably may slightly reduce the Weyl-tensor components, see appendix \ref{m-rotations}); the corresponding orthonormal vectors are obtained by (\ref{e2,e3}).

To summarize the above in an effective way: by a suitable choice of $V^\mu$, it is possible to rotate the eigen-plane of $S^{\mu\nu}$ so as to contain the given null vector $k^\mu$ (the highest-multiplicity PND in our case). The spin structure of our problem has thus been connected with the curvature structure, which is desirable if wishing to discuss the spinning-particle motion in dependence on Petrov type of the background. Namely, the interpretation orthonormal tetrad has thus been chosen in the way described in section \ref{ortho-basis}, while the first vector $k^\mu$ of the associated NP tetrad has been identified with the highest-multiplicity PND, which leads to the respective simplification (vanishing) of some of the Weyl-tensor projections. Were it possible to also identify the second NP-tetrad vector $l^\mu$ with some of the other PNDs, some further Weyl scalars could be made vanish (in particular, in type D it is very advantageous to align $l^\mu$ with the second existing double-degenerate PND), but this is typically not the case, because the $(k^\mu,l^\mu)$ plane has no reason to coincide with the eigen-plane of $S^{\mu\nu}$. Practically, by a suitable choice of $V^\mu$ one can make the eigen-plane of $S^{\mu\nu}$ intersect the $(k^\mu,l^\mu)$ plane right along $k^\mu$, but the other generator $s^\mu$ is thus fixed and cannot in general be rotated to fall in the $(k^\mu,l^\mu)$ plane too.

Now it is clear how the equations (\ref{V,dotp})--(\ref{e3,dotp}) simplify in specific algebraic types. If all the Weyl-tensor PNDs are distinct, the space-time is of Petrov type I. If the null tetrad is chosen so that one of these directions coincides with $k^\mu$, the corresponding scalar $\Psi_0$ vanishes (if the PND was aligned with $l^\mu$, the last scalar $\Psi_4$ would vanish instead). This however does not affect equations (\ref{V,dotp})--(\ref{e3,dotp}) since they lack $\Psi_0$ and $\Psi_4$ ``a priori".
In algebraically special cases, the PND $k^\mu$ is degenerate and further Weyl scalars vanish besides $\Psi_0$: in type II, $k^\mu$ is double and $\Psi_0=\Psi_1=0$ and in type III it is triple and $\Psi_0=\Psi_1=\Psi_2=0$, with obvious effect on equations (\ref{V,dotp})--(\ref{e3,dotp}). We see, however, that even in type-III background, where only $\Psi_3$ survives in the equations, all the force components are still non-zero in general, just one of the components of $\Psi_3$ can be transformed out in addition by a suitable rotation of $m^\mu$ and $\bar{m}^\mu$ vectors (appendix \ref{m-rotations}).

\subsection{Type-N and type-D fields}

The remaining Petrov types N and D are usually discussed separately, since they are algebraically the most special and cover many of important exact space-times known. In type N, corresponding to purely transverse plane waves, there is just one, four times repeated PND. Using it as $k^\mu$ and erecting the orthonormal tetrad as described in section \ref{tetrad-construction}, one obtains equations with $(\Psi_0=)\,\Psi_1=\Psi_2=\Psi_3=0$, hence only keeping the cosmological terms,
\begin{eqnarray}
 -V_\mu\,\dot{p}^\mu=0, &\quad&
  e^{\hat{1}}_\mu\,\dot{p}^\mu=0, \\
  e^{\hat{2}}_\mu\,\dot{p}^\mu
     = -\frac{\Lambda}{3}\,s\,u^{\hat{3}}, &\quad&
  e^{\hat{3}}_\mu\,\dot{p}^\mu
     = \frac{\Lambda}{3}\,s\,u^{\hat{2}}.
\end{eqnarray}
Recalling that $u_\mu\dot{p}^\mu=0$, we see that {\em three} (generically independent) projections of $\dot{p}^\mu$ vanish, so apparently it is possible to rotate $e_{\hat{2}}^\mu$ and $e_{\hat{3}}^\mu$ so that to annul one of the respective projections additionally.
In the ``intrinsic" tetrad suggested in section \ref{intrinsic-tetrad}, the third of these equations writes
\begin{equation}
  {\cal M}\dot{\cal M}=-\frac{\Lambda}{3}\,s\dot{s}
\end{equation}
and the last one, thanks to the second ($s_\mu\dot{p}^\mu=0$), writes
\begin{equation}
  \dot{V}_\mu\dot{p}^\mu=\frac{\Lambda}{3}\,(\gamma m-\mu) \,.
\end{equation}

Finally, in space-times of Petrov type D, there are two independent double PNDs. Identifying again one of them with the NP-tetrad vector $k^\mu$, one arrives at the same result as for type II, namely $(\Psi_0=)\,\Psi_1=0$. A further simplification, $\Psi_3=\Psi_4=0$, only occurs if the second NP-tetrad vector $l^\mu$ can be aligned with the second double PND (this choice is known as the Kinnersley tetrad); all the tidal field is represented then by $\Psi_2$, so the equations reduce to
\begin{eqnarray}
  -V_\mu\,\dot{p}^\mu
     &=& -2s\,{\rm Im}\Psi_2\,u^{\hat{1}}, \label{V,dotp;D} \\
  e^{\hat{1}}_\mu\,\dot{p}^\mu
     &=& -2s\,{\rm Im}\Psi_2\,u^{\hat{0}}, \\
  e^{\hat{2}}_\mu\,\dot{p}^\mu
     &=& +s\left(2\,{\rm Re}\Psi_2-\frac{\Lambda}{3}\right)u^{\hat{3}}, \\
  e^{\hat{3}}_\mu\,\dot{p}^\mu
     &=& -s\left(2\,{\rm Re}\Psi_2-\frac{\Lambda}{3}\right)u^{\hat{2}}. \label{e3,dotp;D}
\end{eqnarray}
In our ``intrinsic" tetrad the third of these equations writes
\begin{equation}
  {\cal M}\dot{\cal M}=\left(2\,{\rm Re}\Psi_2-\frac{\Lambda}{3}\right)s\dot{s} \;.
\end{equation}
As pointed out already, the Kinnersley-like choice is however not feasible in general, if we need to reconcile the tetrad with the spin structure at the same time.

\subsection{Ideally rotated NP tetrads: transverse frames}
\label{transverse}

It is known \cite{BeetleB-02,ReBMW-03} that instead of canceling out the scalars $\Psi_0$ and $\Psi_4$, it is quite generally possible to do the same with the couple $\Psi_1$ and $\Psi_3$. This is in fact a preferable alternative, as it means elimination of the ``pure-gauge" longitudinal-wave effects. Notably, \cite{FerrandoS-05} presented a covariant procedure, applicable to any type-I space-time and any initially chosen NP tetrad, how to find a new tetrad (the {\em transverse frame}) in which $\Psi_1=0$ and $\Psi_3=0$, together with prescriptions for the new values of $\Psi_0$, $\Psi_2$ and $\Psi_4$. Such an option seems ideal for our projections (\ref{V,dotp})--(\ref{e3,dotp}) of the MPD equations, since they do not contain $\Psi_0$ and $\Psi_4$ from the beginning, so their values are irrelevant. Hence, if we used in our picture as $k^\mu$ and $l^\mu$ the vectors reached in Corollary 2 of \cite{FerrandoS-05} (and derive $V^\mu$, $e_{\hat{\imath}}^\mu$ from them accordingly), the MPD-equation projections (\ref{V,dotp})--(\ref{e3,dotp}) would reduce to the type-D form (\ref{V,dotp;D})--(\ref{e3,dotp;D}) in {\em any} field. The only exception is type III where the ``transverse" frame cannot be found; the existence and (non-)uniqueness of such a tetrad were summarized in \cite{BeetleB-02}. Note also that in type N our MPD-equation projections write even simpler (see previous subsection), so there is no need to use the transverse frame. To sum up, the transverse tetrad could simplify our equations in type-I and type-II space-times, effectively turning them into the type-D form.

There is a problem, however:
as found in \cite{FerrandoS-05}, there generally exist {\em three} distinct ``principal" transverse frames $(k^\mu,l^\mu,m^\mu,\bar{m}^\mu)$, plus a continuous set of their derivatives obtained by renormalization $e^\phi k^\mu$, $e^{-\phi}l^\mu$ in the real time-like plane and rotation $e^{-{\rm i}\theta}m^\mu$, $e^{{\rm i}\theta}\bar{m}^\mu$ in the complex space-like plane. Hence, for any of these alternatives, the $(k^\mu,l^\mu)$ plane is {\em fixed}, so -- like with the Kinnersley-tetrad choice in the type-D case -- it is not in general possible to fix it to the spin direction concurrently.

In type-D space-times, ${\rm Re}\Psi_2$ and ${\rm Im}\Psi_2$ represent respectively components of the gravitoelectric and gravitomagnetic tidal fields (appendix \ref{Weyl-electric,magnetic}); ${\rm Re}\Psi_2$ stands for expansion and ${\rm Im}\Psi_2$ stands for vorticity/twist of the PNDs. Clearly ${\rm Re}\Psi_2$ is connected with the ``scalar", centrally acting field component, while ${\rm Im}\Psi_2$ is connected with magnetic-type effects due to mass currents (typically due to rotation). Equations (\ref{V,dotp;D})--(\ref{e3,dotp;D}) reveal that in case one could make the interpretation tetrad transverse (thus definitely not in space-times of type III), the electric part of curvature would drive the spinning particle within the plane orthogonal to $(V^\mu,s^\mu)$ (i.e. within the blade of $S^{\mu\nu}$), while the magnetic part of curvature would drive it within the $(V^\mu,s^\mu)$ plane (i.e. within the blade of ${^*\!}S^{\mu\nu}$).

In type-III space-times where the NP tetrad cannot be rotated so as to become transverse (not even at one point), equations of motion (\ref{V,dotp})--(\ref{e3,dotp}) read
\begin{eqnarray}
  -V_\mu\,\dot{p}^\mu
     &=& -s\,{\rm Im}\Psi_3\,u^{\hat{2}}-s\,{\rm Re}\Psi_3\,u^{\hat{3}} \,, \\
  e^{\hat{1}}_\mu\,\dot{p}^\mu
     &=& -s\,{\rm Im}\Psi_3\,u^{\hat{2}}-s\,{\rm Re}\Psi_3\,u^{\hat{3}} \,, \\
  e^{\hat{2}}_\mu\,\dot{p}^\mu
     &=& -\frac{\Lambda}{3}\,s\,u^{\hat{3}}
         +s\,{\rm Im}\Psi_3\,(u^{\hat{1}}-u^{\hat{0}}) \,, \\
  e^{\hat{3}}_\mu\,\dot{p}^\mu
     &=& +\frac{\Lambda}{3}\,s\,u^{\hat{2}}
         +s\,{\rm Re}\Psi_3\,(u^{\hat{1}}-u^{\hat{0}}) \,.
\end{eqnarray}
The first two projections have the same r.h. side, and the last two would be just Im and Re parts of the same expression if $\Lambda$ were zero.

Note, finally, that the transverse tetrad is (or would be) an {\em alternative} to the ``intrinsic" tetrad suggested in section \ref{intrinsic-tetrad}: when using the intrinsic orthonormal tetrad, given by the pole-dipole description itself (provided that the reference observer $V^\mu$ has been fixed), it is generally {\em not} possible to make it transverse with respect to the curvature at the same time.

\section{Specific spin conditions}
\label{SSCs}

Let us check now whether some of the usually posed spin supplementary conditions bring some advantages when employed in particular Petrov types. It must be stressed, however, that choosing a certain particular $V^\mu$ {\em irrespectively of} the curvature type (typically as fixed to some important direction given by the particle motion) means that this zeroth tetrad vector can{\em not} in general be {\em at the same time} aligned with the PNDs in the desirable way, so in special Petrov types the MPD equation only gets further simplified if the highest-multiplicity PND $k^\mu$ {\em incidentally} belongs to the (now {\em a priori} selected) spin plane $(V^\mu,s^\mu)$.

\subsection{Mathisson--Pirani spin condition, $V^\mu\equiv u^\mu$}
\label{MP-condition}

With the condition $S^{\mu\sigma}u_\sigma=0$, the first MPD equation (\ref{Dp,s}) becomes
\begin{equation}
  \dot{p}^\mu = {^*\!R^\mu}_{\nu\alpha\beta}u^\nu s^\alpha u^\beta
              \equiv B^\mu_\alpha s^\alpha \,,
\end{equation}
where $B_{\alpha\beta}$ is the gravitomagnetic tidal field (appendix \ref{Weyl-electric,magnetic}).
The r.h. side thus differs from that of the geodesic-deviation equation only in sign and in the Hodge dualization of Riemann. The tangent-vector tetrad components degenerate to
\[u^{\hat{0}}\equiv\gamma\equiv -u_\sigma u^\sigma=1 \,, \quad
  u^{\hat{\imath}}=0,\]
which simplifies the (\ref{V,dotp})--(\ref{e3,dotp}) projections to
\begin{eqnarray}
  e^{\hat{1}}_\mu\,\dot{p}^\mu
     &=& -2s\,{\rm Im}\Psi_2,  \label{V,dotp;MP} \\
  e^{\hat{2}}_\mu\,\dot{p}^\mu
     &=& -s\,({\rm Im}\Psi_3-{\rm Im}\Psi_1), \\
  e^{\hat{3}}_\mu\,\dot{p}^\mu
     &=& -s\,({\rm Re}\Psi_3+{\rm Re}\Psi_1)  \label{e3,dotp;MP}
\end{eqnarray}
(the time component is trivial, since $u_\mu\dot{p}^\mu=0$).
Here the cosmological constant dropped out completely, while all three Weyl scalars remain present; $\Psi_2$ is only represented by its imaginary part and determines the force which acts on the particle in the direction of its spin.

For $V^\mu\equiv u^\mu$, one has $\gamma\equiv 1$ and $\mu\equiv m$, equation (\ref{dotm}) implies $\dot{m}=-\dot{u}_\mu p^\mu=0$, equation (\ref{dots}) reduces to $\dot{s}^\mu=u^\mu\dot{u}_\nu s^\nu$, equation (\ref{sds=EsVup}) gives $\dot{s}=0$, and equations (\ref{ps=us}) and (\ref{pdots=udots}) imply
\begin{equation}
  s_\mu u^\mu=0 \;\Rightarrow\; s_\mu p^\mu=0
  \quad (\Rightarrow) \quad
  s_\mu\dot{p}^\mu=ms_\mu\dot{u}^\mu \,,
\end{equation}
which allows to rewrite the first of the above equations (for the spin-direction projection of $\dot{p}^\mu$) directly in terms of four-acceleration $\dot{u}^\mu$,
\begin{equation}
  ms_\mu\dot{u}^\mu\,(=-mu_\mu\dot{s}^\mu)=-2s^2\,{\rm Im}\Psi_2 \,.
\end{equation}
A propos, in the case of the Mathisson--Pirani condition, the four-acceleration can be isolated from the MPD system \cite{KyrianS-07},
\begin{equation}  \label{dotu,MP}
 \dot{u}^\mu
 =\frac{1}{s^2}
 \left(\frac{1}{m}\,\dot{p}^\iota s_\iota s^\mu-p_\kappa S^{\mu\kappa}\right),
\end{equation}
as also following by equation (\ref{dotV}).

As notified above, after choosing $V^\mu\equiv u^\mu$ the tetrad zeroth and first vectors are fixed, so even if the space-time was algebraically special, they cannot be rotated to make $k^\mu\equiv\frac{1}{\sqrt{2}}\,(V^\mu+s^\mu/s)$ coincide with the desired PND. Only in the special case when the particle is moving so that
$k^\mu\equiv\frac{1}{\sqrt{2}}\,(u^\mu+s^\mu/s)$ points just in that principal direction, the $\Psi_1$ scalar vanishes and the above equations simplify accordingly. If the principal direction was even triple degenerate (type-III field) in that case, $\Psi_2$ would vanish too, implying no force (and no acceleration) in the direction of the spin vector. Finally, for a particle moving in the direction given by PND of the type-N space-time, one has $e^{\hat\alpha}_\mu\,\dot{p}^\mu=0$, so $\dot{p}^\mu=0$ (no force).

\subsection{Tulczyjew spin condition, $V^\mu\equiv p^\mu/{\cal M}$}
\label{T-condition}

With $V^\mu\equiv p^\mu/{\cal M}$, one has $\gamma\equiv m/{\cal M}$ and $\mu\equiv{\cal M}$, equation (\ref{dotM}) implies ${\cal M}\dot{\cal M}=-\dot{p}_\mu p^\mu=0$, equation (\ref{dots}) reduces to ${\cal M}^2\dot{s}^\mu=p^\mu\dot{p}_\nu s^\nu$ and equation (\ref{sds=EsVup}) to $\dot{s}=0$, and equations (\ref{ps=us}) and (\ref{pdots=udots}) imply, analogously as above,
\begin{equation}  \label{Tulczyjew-conseq}
  s_\mu p^\mu=0 \;\Rightarrow\; s_\mu u^\mu=0
  \quad (\Rightarrow) \quad
  s_\mu \dot{u}^\mu=\frac{m}{{\cal M}^2}\,s_\mu \dot{p}^\mu \,.
\end{equation}
However, the main advantage of this condition is that an explicit relation exists giving $u^\mu$ in terms of $p^\mu$ and $S^{\mu\nu}$, \cite{Kunzle-72}
\begin{equation}  \label{p-u-relation}
  u^\mu=
   \frac{m}{{\cal M}^2}
   \left(p^\mu+
         \frac{2S^{\mu\nu}R_{\nu\iota\kappa\lambda}p^\iota S^{\kappa\lambda}}
              {4{\cal M}^2+R_{\alpha\beta\gamma\delta}S^{\alpha\beta}S^{\gamma\delta}}
         \right)
\end{equation}
(plus $u_\mu u^\mu=-1$ fixes $m$ which is {\em not} constant along $u^\mu$ here).
Likewise, it is also possible to find $p^\mu$ in terms of $u^\mu$ and $S^{\mu\nu}$. Actually, relation (\ref{ps=us}) implies $u^\mu s_\mu=0$, so $p^\mu$ and $u^\mu$ are both orthogonal to $\dot{p}^\mu$ as well as to $s^\mu$. Hence, it is natural to decompose $p^\mu$ into $mu^\mu$ and a term orthogonal to $u^\mu$ (namely proportional to $\epsilon^{\mu\nu\rho\sigma}u_\nu\dot{p}_\rho s_\sigma$). Multiplying this decomposition by $\epsilon_{\mu\alpha\beta\gamma}u^\alpha\dot{p}^\beta s^\gamma$, using the relation
\begin{equation}  \label{dotp,Sp=0}
  (s^2\delta^\mu_\nu-s^\mu s_\nu)\,\dot{p}^\nu =
  {\cal M}\,\epsilon^{\mu\iota\alpha\beta}s_\iota u_\alpha p_\beta
\end{equation}
following from (\ref{dotV,project}), substituting the definition relation
${\cal M}S^{\mu\alpha}=\epsilon^{\mu\alpha\beta\iota}p_\beta s_\iota$,
and finally demanding that $p_\mu p^\mu=-{\cal M}^2$, one derives
\begin{equation}
  p^\mu=m\,u^\mu-\frac{1}{\cal M}\,\epsilon^{\mu\nu\rho\sigma}u_\nu\dot{p}_\rho s_\sigma \,,
\end{equation}
where $m^2={\cal M}^2+S^{\alpha\beta}\dot{p}_\alpha u_\beta$ and $\dot{p}_\mu$ is to be substituted from the first MPD equation (\ref{Papa-p}).
Note that a counterpart of (\ref{dotp,Sp=0}) can be found as well for the decomposition of $s^\mu$ into components parallel to $\dot{p}^\mu$ and orthogonal to it (thus proportional to $\epsilon^{\mu\nu\rho\sigma}\dot{p}_\nu u_\rho p_\sigma$).

With Tulczyjew's condition, the first MPD equation (\ref{Dp,s}) reads
\begin{equation}  \label{Dp,s;MP}
  {\cal M}\,\dot{p}^\mu = {^*\!R^\mu}_{\nu\alpha\beta}u^\nu s^\alpha p^\beta
\end{equation}
and its ``temporal" ($V_\mu$) projection vanishes again due to
$p_\mu\dot{p}^\mu=-{\cal M}\dot{\cal M}=0$. The tetrad components of four-velocity include
\begin{eqnarray*}
  u^{\hat{0}} &\equiv& \gamma\equiv -\frac{p_\sigma u^\sigma}{\cal M}=\frac{m}{\cal M} \,, \\
  u^{\hat{1}} &:=& \frac{s_\sigma u^\sigma}{s}=\frac{m\,s_\sigma p^\sigma}{{\cal M}^2 s}=0 \,,
\end{eqnarray*}
so the (\ref{V,dotp})--(\ref{e3,dotp}) system reduces to
\begin{eqnarray}
  0=-V_\mu\,\dot{p}^\mu
     &=& +s\,({\rm Im}\Psi_1-{\rm Im}\Psi_3)\,u^{\hat{2}} \nonumber \\
     &{}& -s\,({\rm Re}\Psi_1+{\rm Re}\Psi_3)\,u^{\hat{3}} \,,
     \label{V,dotp;Tulczyjew} \\
  \frac{{\cal M}^2}{m}\,e^{\hat{1}}_\mu\,\dot{u}^\mu=
  e^{\hat{1}}_\mu\,\dot{p}^\mu
     &=& -\frac{2ms}{\cal M}\;{\rm Im}\Psi_2 \nonumber \\
     &{}& -s\,({\rm Im}\Psi_3+{\rm Im}\Psi_1)\,u^{\hat{2}} \nonumber \\
     &{}& -s\,({\rm Re}\Psi_3-{\rm Re}\Psi_1)\,u^{\hat{3}} \,, \\
  e^{\hat{2}}_\mu\,\dot{p}^\mu
     &=& +s\left(2\,{\rm Re}\Psi_2-\frac{\Lambda}{3}\right)u^{\hat{3}} \nonumber \\
     &{}& -\frac{ms}{\cal M}\;({\rm Im}\Psi_3-{\rm Im}\Psi_1) \,, \\
  e^{\hat{3}}_\mu\,\dot{p}^\mu
     &=& -s\left(2\,{\rm Re}\Psi_2-\frac{\Lambda}{3}\right)u^{\hat{2}} \nonumber \\
     &{}& -\frac{ms}{\cal M}\;({\rm Re}\Psi_3+{\rm Re}\Psi_1) \,.
     \label{e3,dotp;Tulczyjew}
\end{eqnarray}
The first of these equations together with normalization
\[-1=-(u^{\hat{0}})^2+\sum(u^{\hat{\imath}})^2
    =-\frac{m^2}{{\cal M}^2}+(u^{\hat{2}})^2+(u^{\hat{3}})^2\]
give $u^{\hat{2}}$ and $u^{\hat{3}}$ in terms of the Weyl scalars,
\begin{eqnarray}
  u^{\hat{2}}
  &=& \pm\,\frac{{\rm Re}\Psi_1+{\rm Re}\Psi_3}{{\cal M}} \;\times \nonumber \\
  & & {} \times
         \sqrt{\frac{m^2-{\cal M}^2}
                    {({\rm Re}\Psi_1+{\rm Re}\Psi_3)^2+({\rm Im}\Psi_1-{\rm Im}\Psi_3)^2}} \nonumber \\
  &=& \pm\,\frac{{\rm Re}\Psi_1+{\rm Re}\Psi_3}{{\cal M}\,|\Psi_1+\overline{\Psi}_3|}
         \,\sqrt{m^2-{\cal M}^2} \;, \\
  u^{\hat{3}}
  &=& \pm\,\frac{{\rm Im}\Psi_1-{\rm Im}\Psi_3}{{\cal M}} \;\times \nonumber \\
  & & {} \times
         \sqrt{\frac{m^2-{\cal M}^2}
                    {({\rm Re}\Psi_1+{\rm Re}\Psi_3)^2+({\rm Im}\Psi_1-{\rm Im}\Psi_3)^2}} \nonumber \\
  &=& \pm\,\frac{{\rm Im}\Psi_1-{\rm Im}\Psi_3}{{\cal M}\,|\Psi_1+\overline{\Psi}_3|}
         \,\sqrt{m^2-{\cal M}^2} \;,
\end{eqnarray}
which can then be used in the remaining equations to express $e^{\hat{\imath}}_\mu\,\dot{p}^\mu$ in terms of $\Psi_a$; for instance,
\begin{equation}  \label{e1,dotp;Tulczyjew}
  e^{\hat{1}}_\mu\,\dot{p}^\mu
     = -\frac{2ms}{\cal M}\;{\rm Im}\Psi_2
       \mp\,\frac{2s\,{\rm Im}\Psi_1\,{\rm Im}\Psi_3}{{\cal M}\,|\Psi_1+\overline{\Psi}_3|}
          \,\sqrt{m^2-{\cal M}^2} \;.
\end{equation}

Again, if $p^\mu$ {\em incidentally} points in such a direction that the plane $(p^\mu,s^\mu)$ contains some of the PNDs of an algebraically special space-time, then $\Psi_1=0$ and
\begin{eqnarray}
  e^{\hat{1}}_\mu\,\dot{p}^\mu
     &=& -\frac{2ms}{\cal M}\;{\rm Im}\Psi_2 \,, \\
  e^{\hat{2}}_\mu\,\dot{p}^\mu
     &=& -\frac{s}{\cal M}\,{\rm Im}\Psi_3
          \left(m\pm\frac{2\,{\rm Re}\Psi_2-\frac{\Lambda}{3}}{|\Psi_3|}
                    \sqrt{m^2-{\cal M}^2}\right), \\
  e^{\hat{3}}_\mu\,\dot{p}^\mu
     &=& -\frac{s}{\cal M}\,{\rm Re}\Psi_3
          \left(m\pm\frac{2\,{\rm Re}\Psi_2-\frac{\Lambda}{3}}{|\Psi_3|}
                    \sqrt{m^2-{\cal M}^2}\right).
\end{eqnarray}
(One of the last two projections can furthermore be transformed out by a suitable rotation of $m^\mu$ and $\bar{m}^\mu$ vectors.)
A similar situation in the type-III case would mean no force in the spin-vector direction (plus one more transformable out), and if it were type N, the force would vanish completely again.

A short remark to equation (\ref{dotp,Sp=0}). It says that projection of $\dot{p}^\mu$ onto a hypersurface orthogonal to $s^\mu\,(\equiv s\,e^\mu_{\hat{1}})$ equals ${\cal M}\,\epsilon^{\mu\iota\alpha\beta}s_\iota u_\alpha p_\beta$, which means that the $e^{\hat{2}}_\mu\,\dot{p}^\mu$ and $e^{\hat{3}}_\mu\,\dot{p}^\mu$ components can also be obtained directly from that expression (rather than from (\ref{Dp,s;MP}) which is more complicated),
\begin{align*}
  s\,e_\mu^{\hat{2}}\,\dot{p}^\mu
    &= \frac{\cal M}{s}\,\epsilon^{\mu\iota\alpha\beta}e_\mu^{\hat{2}}s_\iota u_\alpha p_\beta
     = {\cal M}\,\epsilon^{\hat{2}\hat{1}\hat{\alpha}\hat{\beta}}u_{\hat{\alpha}} p_{\hat{\beta}}
     = \\
    &= {\cal M}\,(u_{\hat{0}}p_{\hat{3}}-u_{\hat{3}}p_{\hat{0}})
     = {\cal M}^2 u^{\hat{3}}-mp^{\hat{3}} \,, \\
  s\,e_\mu^{\hat{3}}\,\dot{p}^\mu
    &= \frac{\cal M}{s}\,\epsilon^{\mu\iota\alpha\beta}e_\mu^{\hat{3}}s_\iota u_\alpha p_\beta
     = {\cal M}\,\epsilon^{\hat{3}\hat{1}\hat{\alpha}\hat{\beta}}u_{\hat{\alpha}} p_{\hat{\beta}}
     = \\
    &= {\cal M}\,(u_{\hat{2}}p_{\hat{0}}-u_{\hat{0}}p_{\hat{2}})
     = mp^{\hat{2}}-{\cal M}^2 u^{\hat{2}} \,.
\end{align*}
However, this is nothing new, it only reproduces what we know from the beginning, namely relation (\ref{u,p,dSp}). Actually, with the Tulczyjew condition, the latter reads
\begin{align*}
  {\cal M}^2 u^\alpha-m\,p^\alpha
    &= -S^{\alpha\beta}\dot{p}_\beta \\
    &= -\epsilon^{\alpha\beta\gamma\delta}V_\beta s_\gamma \dot{p}_\delta
     = s\,\epsilon^{\hat{0}\hat{1}\alpha\delta}\dot{p}_\delta
\end{align*}
which exactly yields the above projections.

\subsubsection{Using ``intrinsic" tetrad with Tulczyjew's condition}

The Tulczyjew condition very well fits together with decomposing the MPD equation in the ``intrinsic" tetrad suggested in section \ref{intrinsic-tetrad}. Actually, substituting $V^\mu=p^\mu/{\cal M}$, $\gamma=m/{\cal M}$, $\mu={\cal M}$ and $\dot{s}=0$ into (\ref{u2,intrinsic})--(\ref{dotp3,intrinsic}), and assuming $m\neq{\cal M}$ (otherwise the tetrad is not defined), we obtain for $u^\mu$ and $\dot{p}^\mu$ the components
\begin{align}
  u^{\hat{2}} &= \frac{\sqrt{m^2-{\cal M}^2}}{\cal M} \;, \\
  u^{\hat{3}} &= \frac{m\,\dot{s}}{{\cal M}\,\sqrt{m^2-{\cal M}^2}} = 0 \,, \\
  e^{\hat{2}}_\mu\,\dot{p}^\mu
     &= \frac{m\,\dot{\cal M}}{\sqrt{m^2-{\cal M}^2}}=0 \,, \\
  e^{\hat{3}}_\mu\,\dot{p}^\mu
     &= \frac{-p^\mu\dot{S}_{\mu\lambda}\dot{S}^{\lambda\nu}p_\nu}{{\cal M}s\,\sqrt{m^2-{\cal M}^2}}
      = \frac{\cal M}{s}\,\sqrt{m^2-{\cal M}^2} \;,
\end{align}
which reduces the (\ref{V,dotp;Tulczyjew})--(\ref{e3,dotp;Tulczyjew}) system to
\begin{eqnarray}
  0 &=& s\,({\rm Im}\Psi_1-{\rm Im}\Psi_3) \,, \label{V,dotp;Tulczyjew,intrinsic} \\
  e^{\hat{1}}_\mu\,\dot{p}^\mu
     &=& -\frac{2ms}{\cal M}\;{\rm Im}\Psi_2
         -s\,({\rm Im}\Psi_1+{\rm Im}\Psi_3)\,u^{\hat{2}} \,, \\
  0 &=& s\,({\rm Im}\Psi_1-{\rm Im}\Psi_3) \,,
     \label{e2,dotp;Tulczyjew,intrinsic} \\
  {\cal M}^2 &=& s^2\!\left(\!\frac{\Lambda}{3}\!-\!2\,{\rm Re}\Psi_2\!\right)
                 -\frac{ms^2\,({\rm Re}\Psi_1\!+\!{\rm Re}\Psi_3)}{\sqrt{m^2-{\cal M}^2}} \,.
     \label{e3,dotp;Tulczyjew,intrinsic}
\end{eqnarray}
The first and the third equation are equal and imply ${\rm Im}\Psi_1={\rm Im}\Psi_3$; the last equation represents a {\em constraint} between several parameters of the exercise; evolution has only remained in the $s_\mu\dot{p}^\mu$ component.

If the $(p^\mu,s^\mu)$ plane contained some of the PNDs of an algebraically special field, one would have $\Psi_1=0$, hence also ${\rm Im}\Psi_3=0$, and the evolution equation would shorten to
\[\left(\frac{{\cal M}^2}{m}\,e^{\hat{1}}_\mu\,\dot{u}^\mu=\right)
  e^{\hat{1}}_\mu\,\dot{p}^\mu=-\frac{2ms}{\cal M}\;{\rm Im}\Psi_2 \,.\]
In type III where $\Psi_2=0$ as well, this r.h. side would vanish and the constraint would reduce to
\[{\cal M}^2 = \frac{\Lambda}{3}\,s^2
               -\frac{ms^2\,{\rm Re}\Psi_3}{\sqrt{m^2-{\cal M}^2}} \;.\]
We have not mentioned types D and N, because in these cases the ``intrinsic" tetrad is not available.
To prove this, let us evaluate the term which ``deviates" $u^\mu$ from $p^\mu$ according to the relation (\ref{p-u-relation}):
by substituting consecutively
\begin{align*}
  & S^{\kappa\lambda}=-{\epsilon^{\kappa\lambda}}_{\alpha\beta}s^\alpha V^\beta,
    \quad
    p^\iota={\cal M}V^\iota,
    \quad
    S^{\mu\nu}=\epsilon^{\mu\nu\rho\sigma}V_\rho s_\sigma, \\
  & R^*_{\nu\hat{0}\hat{1}\hat{0}}={^*\!}R_{\nu\hat{0}\hat{1}\hat{0}}={^*\!}C_{\nu\hat{0}\hat{1}\hat{0}}
    \quad ({\rm vacuum}),
\end{align*}
decomposing the result in the generic $(k^\mu,l^\mu,m^\mu,\bar{m}^\mu)$-tied orthonormal tetrad and using Appendix \ref{Appendix-A}, one finds
\begin{align*}
  &S^{\mu\nu}R_{\nu\iota\kappa\lambda}p^\iota S^{\kappa\lambda}
   =-2S^{\mu\nu}R^*_{\nu\iota\alpha\beta}p^\iota s^\alpha V^\beta= \\
  &=-2{\cal M}s\,S^{\mu\nu}\,{^*\!}R_{\nu\hat{0}\hat{1}\hat{0}}
   =-2{\cal M}s^2\,\epsilon^{\mu\nu\hat{0}\hat{1}}\,{^*\!}C_{\nu\hat{0}\hat{1}\hat{0}}= \\
  &=2{\cal M}s^2\left({^*\!}C_{\hat{3}\hat{0}\hat{1}\hat{0}}e_{\hat{2}}^\mu-
                      {^*\!}C_{\hat{2}\hat{0}\hat{1}\hat{0}}e_{\hat{3}}^\mu\right)= \\
  &=-2{\cal M}s^2\left[({\rm Re}\Psi_1\!+\!{\rm Re}\Psi_3)\,e_{\hat{2}}^\mu+
                       ({\rm Im}\Psi_1\!-\!{\rm Im}\Psi_3)\,e_{\hat{3}}^\mu\right].
\end{align*}
Hence, in type-D and type-N space-times where $\Psi_1=0=\Psi_3$, this term vanishes and $u^\mu$ is parallel with $p^\mu$, which is exactly the situation when the ``intrinsic" tetrad does not exist.

\subsubsection{Note on the momentum-velocity relation}

As seen from (\ref{tilde(p)}), for the momentum-velocity relation (\ref{u-p,Puetzfeld}) to ``close", a condition weaker than the Tulczyjew's one is in fact sufficient, namely $\frac{{\rm D}}{{\rm d}\tau}(S^{\alpha\beta}p_\beta)$ has to be expressible in terms of $S^{\alpha\beta}$ and $p^\alpha$ only (for example, be proportional to $S^{\alpha\beta}p_\beta$ or to $p^\alpha$). Concerning the importance of the momentum-velocity relation, it might be interesting to examine the range of this option, but we will not go in this direction here.

\subsection{$V^\mu$ parallel along $u^\mu$}
\label{our-condition}

We suggested in \cite{KyrianS-07} to take $V^\mu$ given by some vector parallel along $u^\mu$, i.e. satisfying $\dot{V}^\mu\!=\!0$. Relation (\ref{p,u,SdV}) then implies $u^\mu\parallel p^\mu$ (such an option was already recommended by \cite{Ohashi-03}), therefore $m={\cal M}$, $\mu=\gamma m$ and $\dot{S}^{\mu\nu}\!=\!0$, ${^*\!}\dot{S}_{\alpha\beta}\!=\!0$, $\dot{s}=0$. The mass $m\!=\!{\cal M}$ is constant along $p^\mu=mu^\mu$ and the MPD-equation l.h. side can thus be written as $\dot{p}^\mu={\cal M}\dot{u}^\mu$; its ``time" component, in particular, also reads $-V_\mu\dot{p}^\mu=\dot{\mu}=\dot{\gamma}m$. From $u^\mu\parallel p^\mu$ and $\dot{V}^\mu=0$ it also follows that $\dot{s}^\mu\!=\!0$. Most of the equations in section \ref{MPD-equations} become trivial.

Not so the first MPD equation (\ref{Dp,s}). Actually, in spite of these significant simplifications, the scheme (\ref{V,dotp})--(\ref{e3,dotp}) does {\em not} reduce in general. In particular, note that although the present spin condition leaves the reference vector $V^\mu$ more free, namely only restricted by $\dot{V}^\mu\!=\!0$, one still cannot count with correlating it with the main PND $k^\mu$ besides: in such a case $V^\mu$ and the corresponding $s^\mu$ would have to be related to $k^\mu$ by $V^\mu=\sqrt{2}\,k^\mu-s^\mu/s$, so the requirement $\dot{V}^\mu=0$ would only be fulfilled if $\dot{k}^\mu=0$ (because $\dot{s}^\mu\!=\!0$), i.e. only if $k^\mu$ were itself parallel along the particle's world-line. Of course, it is not in general. On the other hand, if restricting to purely {\em local} analysis, ``at any single point" of the trajectory, then it is always possible to select $V^\mu$ in the desirable way, namely to take advantage of its freedom and choose it {\em there} in the same way as described in section \ref{tetrad-construction}. Therefore, at any given point, we can keep the recipe from that section and simplify equations in some of the algebraically special situations accordingly.

\subsubsection{More on the $u^\mu\parallel p^\mu$ option}
\label{our-condition-generalized}

The main benefit of choosing such $V^\mu$ whose $\dot{V}^\mu=0$ is that $u^\mu$ and $p^\mu$ are parallel then, $p^\mu=m u^\mu$, i.e. the ``hidden" component of momentum (\ref{hidden-p}) vanishes. Besides obvious simplification, this circumstance also remedies one of inherent inconveniences of the extended-body problem. Namely, even though $p^\mu$ should be time-like in reality (i.e., one supposes ${\cal M}^2>0$), the MPD equations do {\em not} in general guarantee that $u^\mu$ is time-like as well (the selected representative world-line may be winding through the body's convex-hull world-tube in an awkward way); in particular, $u^\mu$ has been observed to easily become space-like in highly non-homogeneous fields where the pole-dipole approximation is most problematic. The $u^\mu\parallel p^\mu$ option thus eliminates the need to control the space-time character of $u^\mu$, this being fixed by the character of $p^\mu$. Moreover, since $u^\mu\parallel p^\mu$ implies $\dot{\cal M}=\dot{m}=0$, the normalization of $p^\mu=mu^\mu$ is {\em conserved}. The same of course applies to the reference vector function $V^\mu$, since it is parallel transported along $u^\mu$.

Further advantage of having $p^\mu=m u^\mu$ is that it can keep the problem linear in spin. Actually, the MPD equations (\ref{Papa-p})--(\ref{Papa-S}) themselves {\em are} linear in $S^{\mu\nu}$, but the spin supplementary condition -- which anyway has to be added -- brings the non-linearity in general. The non-linearity can be seen as entering through the momentum-velocity relation which has to be used on the left-hand side of (\ref{Papa-p}) in order to write the latter down as an equation for $u^\mu$; the momentum-velocity relation (\ref{p-u-relation}) arising for the Tulczyjew spin condition is a clear example. With the option $p^\mu=m u^\mu$ (and $m$ constant along $u^\mu$), equations (\ref{Papa-p}) and (\ref{Dp,s}) simply become
\begin{equation}
  m\dot{u}^\mu
  = -\frac{1}{2}\,{R^\mu}_{\nu\kappa\lambda}u^\nu S^{\kappa\lambda}
  = {^*\!R^\mu}_{\nu\alpha\beta}u^\nu s^\alpha V^\beta.
\end{equation}
Therefore, the problem remains linear in spin (if $V^\mu$ is not spin-dependent, of course).

Let us realize now that $u^\mu\parallel p^\mu$ can however be ensured by a weaker condition than $\dot{V}^\mu=0$, namely, it is sufficient to take $\dot{V}^\mu$ proportional to $s^\mu$. Actually, we know that $u^\mu\parallel p^\mu$ implies $\dot{S}^{\alpha\beta}=0$, ${^*\!}\dot{S}^{\alpha\beta}=0$ and $\dot{s}=0$, irrespectively of the spin supplementary condition. But since {\em some} spin condition ($V_\mu S^{\mu\nu}=0$) has ultimately to be employed, it is reasonable to rewrite the $\dot{S}^{\alpha\beta}=0$ option as $\dot{V}_\mu S^{\mu\nu}=0$, as also given by (\ref{p,u,SdV}). Well, we have in fact only restored the statement that the hidden momentum (\ref{hidden-p}) should be zero. Or, in still other words, $u^\mu\parallel p^\mu$ $\Leftrightarrow$ $\dot{V}^\mu$ belongs to the eigen-plane of $S^{\mu\nu}$. The latter is spanned by $V^\mu$ and $s^\mu$, of which $V^\mu$ is nevertheless {\em perpendicular} to $\dot{V}^\mu$, so $\dot{V}^\mu$ has to be proportional to $s^\mu$. Multiplying this proportionality by $s_\mu$, one finds
\begin{equation}  \label{dotV,u||p}
  \dot{V}^\mu=\frac{1}{s^2}\,\dot{V}^\nu s_\nu\,s^\mu.
\end{equation}
This result is nothing new, we know it from equation (\ref{dotV,project}) already.
Similarly, the spin-vector evolution (\ref{dots}) for $u^\mu\parallel p^\mu$ reduces to
\begin{equation}  \label{dots,u||p}
  \dot{s}^\mu=\dot{V}^\nu s_\nu\,V^\mu \,.
\end{equation}

The reference vector function $V^\mu$ is clearly not fixed uniquely. Specifically, it is constrained by $V_\mu V^\mu=-1$ and (\ref{dotV,u||p}), of which the latter represents just two independent conditions, because its projections onto $V^\mu$ and $s^\mu$ are satisfied automatically. The remaining indeterminacy of $V^\mu$ can be interpreted as a freedom to choose the magnitude of $\dot{V}^\mu$. If $\dot{V}^\mu$ is multiplied by some scalar, $\dot{s}_\mu$ has to be multiplied by the same one, otherwise $\frac{{\rm D}}{{\rm d}\tau}(V^\mu s_\mu)$ (and ergo also $V^\mu s_\mu$ itself) would not stay zero.

Consider now how to exploit the above freedom in order to choose $V^\mu$ in a ``natural" way. One possibility is to require $\frac{{\rm D}}{{\rm d}\tau}(u^\mu s_\mu)=0$ which, according to relation (\ref{ps=us}) (but here even more so due to $u^\mu\parallel p^\mu$) is equivalent to
\[\frac{{\rm D}}{{\rm d}\tau}(p^\mu s_\mu)=\dot{p}^\mu s_\mu+p^\mu\dot{s}_\mu=0 \,.\]
Taking now $\dot{s}_\mu\,(=\!\dot{V}^\nu s_\nu V_\mu)=\!\frac{\alpha\,s^2}{\mu{\cal M}^2}\,V_\mu$, with $\alpha$ some dimensionless scalar and $\mu\!\equiv\!-V_\sigma p^\sigma$, ${\cal M}^2\!\equiv\!-p_\sigma p^\sigma$ ($=\!m^2$ when $u^\mu\parallel p^\mu$), we find
\begin{align}
  \alpha=\frac{{\cal M}^2}{s^2}\,\dot{p}^\mu s_\mu
        & =\frac{m^2}{s^2}\,{^*\!R}_{\mu\nu\alpha\beta}s^\mu u^\nu s^\alpha V^\beta \\
        & =m^2\,{^*\!R}_{\hat{1}\hat{\gamma}\hat{1}\hat{0}}u^{\hat{\gamma}} \,,
\end{align}
which can in general (general {\em vacuum}) be decomposed by equation (\ref{e1,dotp}).\footnote
{Specifically in type-N space-times, $\alpha=0$.}
Therefore, if $p^\mu s_\mu=mu^\mu s_\mu=0$ at some (initial) point and $\alpha$ is chosen as above, then $u^\mu s_\mu=0$ ($s^\mu$ is ``purely spatial") along the whole representative world-line.

Finally, a natural option how to set $u^\mu s_\mu=0$ is to simply select $V^\mu\equiv u^\mu$ $(=p^\mu/m)$ initially.
With $\alpha$ chosen by the above prescription, and with $V^\mu$ and $s^\mu$ evolving according to
\begin{equation}
  \dot{V}^\mu=\frac{\alpha}{\mu\,m^2}\,s^\mu,
  \qquad
  \dot{s}^\mu=\frac{\alpha\,s^2}{\mu\,m^2}\,V^\mu,
\end{equation}
the four-momentum $p^\mu$ will then remain tangent and the spin $s^\mu$ orthogonal to the representative world-line $u^\mu$.

\section{Special types of motion}
\label{special-motions}

It is useful to once more realize what can actually be chosen freely in the spinning-particle exercise. Tackling it as a 3+1 problem (e.g. when integrating the MPD equations on computer), one typically first selects $V^\mu$ and the three-vectors of initial relative velocity and initial spin with respect to some observer (which may be different from $V^\mu$); these determine the initial four-velocity $u^\mu$ and four-spin $s^\mu$; then the initial bivector $S_{\alpha\beta}=\epsilon_{\alpha\beta\gamma\delta}V^\gamma s^\delta$ is calculated. The remaining point is to obtain the initial four-momentum $p^\mu$; this is practically done in dependence on the chosen spin supplementary condition, but {\em in principle} $p^\mu$ follows by integrating the energy-momentum tensor over a hypersurface fixed by $V^\mu$. Hence, apart from initial conditions, $V^\mu$ is {\em the only} freely selectable quantity.

It is thus natural that we have first considered the choice of $V^\mu$, because this may be done without loss of generality. In this section, secondarily, let us check whether some ``clean" cases do not follow for special types of motion, i.e. for special $u^\mu$ or/and $p^\mu$ (whether with special choice of $V^\mu$ or not). Note that some of these have already been mentioned within previous section on specific spin conditions.

It should be emphasized that one must distinguish between special setting holding {\em at one point} and the much stronger (and by default considered) circumstance of such a setting remaining valid {\em along the whole representative world-line}.

\subsection{Special $u^\mu$}

The MPD components (\ref{V,dotp})--(\ref{e3,dotp}) simplify when some of the four-velocity components $u^{\hat{\alpha}}$ vanish.

\subsubsection{$u^\mu$ lying in the Weyl-tensor eigen-plane}

If $u^\mu$ lies in the plane spanned by $k^\mu$ and $l^\mu$, it is orthogonal to $m^\mu$ and $\bar{m}^\mu$, hence $u^{\hat{2}}=0$ and $u^{\hat{3}}=0$, with obvious effect on the above equations. However, the plane of $k^\mu$ and $l^\mu$ is the plane of $V^\mu$ and $s^\mu$, namely it is the eigen-plane of $S^{\mu\nu}$, hence necessarily $S^{\mu\sigma}u_\sigma=0$. If the above holds along the trajectory (not just at one point), one is thus back at Mathisson--Pirani condition, section \ref{MP-condition}.

\subsubsection{$u^\mu$ (and thus $p^\mu$) orthogonal to $s^\mu$}

If $u^\mu$ is perpendicular to $s^\mu$ or, in other words, $u^\mu$ is tangent to a time-like hypersurface spanned by $V^\mu$, $m^\mu$ and $\bar{m}^\mu$, then $u^{\hat{1}}\!=\!0$ disappears from the equations. If the transverse NP frame could be used, thus having equations in the (\ref{V,dotp;D})--(\ref{e3,dotp;D}) form, the time component would vanish in that case, $-V_\mu\,\dot{p}^\mu=0$.

Let us check for further consequences, mainly for how the choice of $V^\mu$ is restricted by the requirement $s^\alpha u_\alpha=0$. Firstly, relation (\ref{ps=us}) says that $s^\alpha p_\alpha=0$ then, too. We know that such a situation can be accomplished by selecting $V^\mu=u^\mu$ or $V^\mu=p^\mu/{\cal M}$, i.e. by Mathisson--Pirani or Tulczyjew choice of the spin supplementary condition (sections \ref{MP-condition} and \ref{T-condition} above), and that the simultaneous orthogonality of $s^\mu$ to both $u^\mu$ and $p^\mu$ may also happen when these two vectors are parallel (section \ref{our-condition}). However, here we want to check whether there are some other alternatives, so we assume that the three time-direction vectors $V^\mu$, $u^\mu$ and $p^\mu$ are independent.

Now, since $s^\mu$ is orthogonal to all of them, it can be written
\begin{equation}  \label{s,ortho}
  s^\mu = \frac{s}{\dot{s}}\;\epsilon^{\mu\iota\kappa\lambda}V_\iota u_\kappa p_\lambda \,,
\end{equation}
where (\ref{sds=EsVup}) has been used for ``normalization".
Consequently, equation (\ref{dots}) assumes the form
\begin{equation}  \label{dots,ortho}
  \dot{s}^\mu=V^\mu\dot{V}_\nu s^\nu+\frac{\dot{s}}{s}\,s^\mu \,.
\end{equation}

Does anything follow for the evolution of $V^\mu$?
Projecting the equation (\ref{dotV}) on $u^\mu$ and $p^\mu$, one finds respectively
\begin{eqnarray}
  u_\mu\,\frac{{\rm D}}{{\rm d}\tau}(s V^\mu)
  \equiv u_\mu(s\dot{V}^\mu+\dot{s}V^\mu) &=& 0 \,, \\
  p_\mu\,\frac{{\rm D}}{{\rm d}\tau}(s V^\mu)
  \equiv p_\mu(s\dot{V}^\mu+\dot{s}V^\mu) &=& 0 \,,
\end{eqnarray}
which means orthogonality to both $u^\mu$ and $p^\mu$, because the vector $(s\dot{V}^\mu+\dot{s}V^\mu)$ cannot be trivial. Hence, this vector has to be an eigen-vector of $\dot{S}^{\mu\nu}$ (with zero eigen-value again). It is also simple to check, e.g. by multiplying relation (\ref{dots,ortho}) by $\dot{V}_\mu$, that
\begin{equation}
  \dot{s}_\mu(s\dot{V}^\mu+\dot{s}V^\mu) = 0 \,.
\end{equation}
Since the vectors $u^\mu$, $p^\mu$ and $\dot{s}^\mu$ have to be independent,\footnote
{Otherwise (\ref{dots,ortho}) would be a combination of $u^\mu$ and $p^\mu$, i.e.
\[V^\mu\dot{V}_\nu s^\nu+(\dot{s}/s)\,s^\mu=Au^\mu+Bp^\mu \,.\]
Multiplying this by $s_\mu$, we find $s\dot{s}=0$, which however also equals $\epsilon^{\mu\nu\alpha\beta}s_\mu V_\nu u_\alpha p_\beta$ according to (\ref{sds=EsVup}). Now, $u^\mu$, $p^\mu$ and $V^\mu$ are assumed to be independent in this part, so $s^\mu$ would have to be dependent, which is in contradiction with its being orthogonal to all the three.}
this means that $\frac{{\rm D}}{{\rm d}\tau}(sV^\mu)$ is orthogonal to all $u^\mu$, $p^\mu$ and $\dot{s}^\mu$, so it can also be represented as
\begin{equation}
  s\dot{V}^\mu+\dot{s}V^\mu
  =\frac{1}{\dot{s}}\,\epsilon^{\mu\iota\kappa\lambda}\dot{s}_\iota u_\kappa p_\lambda
  =-\frac{1}{\dot{s}}\,{^*\!}\dot{S}^{\mu\iota}\dot{s}_\iota
\end{equation}
(cf. its generic decomposition (\ref{dotV})).
Besides $\frac{{\rm D}}{{\rm d}\tau}(sV^\mu)$, the other eigen-vector of $\dot{S}^{\mu\nu}$ is of course $s^\mu$, its derivative $\dot{s}^\mu$ belonging to the eigen-plane of $S^{\mu\nu}$ conversely.\footnote
{The second eigen-vector of $\dot{S}^{\mu\nu}$ {\em orthogonal} to $s^\mu$ is $\epsilon^{\mu\iota\alpha\beta}s_\iota u_\alpha p_\beta$. The vector $(s\dot{V}^\mu+\dot{s}V^\mu)$ was already decomposed into this basis in equation (\ref{dotV}).}

Before continuing, two points should be stressed:
\begin{itemize}
\item
It might seem that $\dot{V}^\mu$ is aligned with $s^\mu$ (which would ensure $u^\mu\parallel p^\mu$), because both these vectors are orthogonal to the triple $V^\mu$, $(\mu u^\mu-\gamma p^\mu)$ and $(s\dot{s}^\mu-\dot{s}s^\mu)$ (for $s^\mu$ it is {\em always} so, while $\dot{V}^\mu$ is only orthogonal to the first two in general); note that the second vector of the triple is orthogonal to both the remaining two. But the suspicion is not the case, because exactly in case when $s^\mu$ is orthogonal to $u^\mu$ and $p^\mu$, the vector $(s\dot{s}^\mu-\dot{s}s^\mu)$ is proportional to $V^\mu$ (it is clear from equation (\ref{dots,ortho})), so the triple is not independent.
\item
Mind that orthogonality to both $u^\mu$ and $p^\mu$ does {\em not} mean lying in the plane spanned by $V^\mu$ and $s^\mu$: as $V^\mu$, $p^\mu$ and $u^\mu$ are all time-like, they are never orthogonal to each other, so the planes $(V^\mu,s^\mu)$ and $(u^\mu,p^\mu)$ are never orthogonal, in spite of $s^\mu$ being orthogonal to both $u^\mu$ and $p^\mu$ (in this subsection).
\end{itemize}

Let us assume that the orthogonalities $s^\alpha u_\alpha=0$ and $s^\alpha p_\alpha=0$ remain valid all along the representative world-line, i.e. that $\frac{{\rm D}}{{\rm d}\tau}(s^\alpha u_\alpha)=0$, $\frac{{\rm D}}{{\rm d}\tau}(s^\alpha p_\alpha)=0$ as well. Combining these with the relation (\ref{pdots=udots}), i.e. with $\gamma\,p^\alpha\dot{s}_\alpha=\mu\,u^\alpha\dot{s}_\alpha$, we also see that
\begin{equation}
  \gamma\,\dot{p}^\alpha s_\alpha=\mu\,\dot{u}^\alpha s_\alpha \,,
\end{equation}
which can further be extended on account of relation (\ref{dots}):
\begin{eqnarray}
   \gamma\,\dot{p}^\alpha s_\alpha
   &=& \mu\,\dot{u}^\alpha s_\alpha
    =  \gamma\mu\,\dot{V}^\alpha s_\alpha = \nonumber \\
    = -\gamma\,p^\alpha\dot{s}_\alpha
   &=&-\mu\,u^\alpha\dot{s}_\alpha
    = -\gamma\mu\,V^\alpha\dot{s}_\alpha \,.
\end{eqnarray}
The above means that we already know of several vectors orthogonal to $s^\mu$: $V^\mu$, $p^\mu$, $u^\mu$ and $(\gamma\dot{p}^\mu-\mu\dot{u}^\mu)$, of which the last two are also orthogonal to each other. Hence, $(\gamma\dot{p}^\mu-\mu\dot{u}^\mu)$ must be some combination of $V^\mu$, $p^\mu$ and $u^\mu$, because these three are independent by assumption.

\subsubsection{$\dot{u}^\mu=0$: geodesic motion}

It is known that in special situations the spin-curvature interaction may have no effect on the particle's four-velocity, thus leaving the motion free. Vanishing of acceleration $\dot{u}^\mu$ implies, irrespectively of spin condition,
\begin{eqnarray}
  \dot{m} &=& -\dot{u}_\mu p^\mu=0 \,, \\
  \dot{p}^\mu &=& -\ddot{S}^{\mu\sigma}u_\sigma \,.
\end{eqnarray}
Note that $\dot{S}^{\alpha\beta}\dot{u}_\beta=0$ holds due to $\dot{u}_\beta=0$ here, so one cannot argue that $\dot{u}^\mu$ is another eigen-vector of $\dot{S}^{\alpha\beta}$, the latter thus being trivial, etc.

When $\dot{u}^\mu=0$, the Mathisson--Pirani condition $V^\mu=u^\mu$ cleary coincides with the $\dot{V}^\mu=0$ condition. This is an advantageous option, since the latter yields $p^\mu=mu^\mu$ and $\dot{p}^\mu=m\dot{u}^\mu\,(=0)$, so there is no force and the MPD exercise reduces to the constraint
\begin{equation}
  {^*\!R^\mu}_{\nu\alpha\beta}u^\nu s^\alpha u^\beta\equiv B^\mu_\alpha s^\alpha=0 \,,
\end{equation}
or, if written out in terms of its projections (\ref{V,dotp;MP})--(\ref{e3,dotp;MP}),
\begin{equation}
  {\rm Im}\Psi_2=0, \;
  {\rm Im}\Psi_1\!-\!{\rm Im}\Psi_3=0, \;
  {\rm Re}\Psi_1\!+\!{\rm Re}\Psi_3=0 \,.
\end{equation}
Hence, the $B_{\alpha\beta}$ field has to be zero in the rest frame of the particle, or, if this is not the case, the particle's spin has to be its eigen-vector (with zero eigen-value).

On the other hand, the Tulczyjew condition $S^{\alpha\beta}p_\beta=0$ yields different results, in particular, it allows for non-zero force $\dot{p}^\mu$ even when $\dot{u}^\mu=0$. Regarding (\ref{Tulczyjew-conseq}), we however see that $s_\mu\dot{p}^\mu=0$, which in turn reduces the expression (\ref{dotp,Sp=0}) to
\begin{equation}  \label{dotp=epsilon(s,u,p)}
  s^2\dot{p}^\mu = {\cal M}\,\epsilon^{\mu\iota\alpha\beta}s_\iota u_\alpha p_\beta \,.
\end{equation}
Also, using $V_\nu\equiv p_\nu/{\cal M}$ and $\dot{p}_\nu s^\nu=0$ in (\ref{dots}), we have
\begin{equation}
  \dot{s}^\mu=0
  \quad (\Rightarrow \dot{s}=0) \,.
\end{equation}
Finally, since $e^{\hat{1}}_\mu\,\dot{p}^\mu=0$, the Tulczyjew-condition form of the MPD decomposition, (\ref{V,dotp;Tulczyjew})--(\ref{e3,dotp;Tulczyjew}), reduces to the last two projections, plus the condition that the expression (\ref{e1,dotp;Tulczyjew}) has to yield zero. The latter says that the masses $m$ and ${\cal M}$ are necessarily related through the scalars $\Psi_1$, $\Psi_3$ and ${\rm Im}\,\Psi_2$. Equivalently (and more simply), one can calculate $e^{\hat{2}}_\mu\,\dot{p}^\mu$ and $e^{\hat{3}}_\mu\,\dot{p}^\mu$ from (\ref{dotp=epsilon(s,u,p)}). As the latter is clearly space-like and orthogonal to $s^\mu$, it is moreover possible to rotate the $e^\mu_{\hat{2}}$, $e^\mu_{\hat{3}}$ vectors so that it have just a single component.

\subsubsection{Stationary situation}

Consider now a situation when relevant scalars do not change along $u^\mu$. It is in fact sufficient to demand
\[\dot{\cal M}=0 \quad \Rightarrow \quad
                 \dot{S}^{\alpha\beta}\dot{p}_\beta=0, \;\; \dot{p}_\alpha p^\alpha=0 \,,\]
which means that $\dot{p}_\beta$ is an eigen-vector of $\dot{S}^{\alpha\beta}$. Since $\dot{p}_\beta$ (supposed to be non-zero) is independent of the generic eigen-vectors of $\dot{S}^{\alpha\beta}$ given in section \ref{eigen-vectors}, the bivector $\dot{S}^{\alpha\beta}$ has to be zero-rank (antisymmetry only allows for even rank, so it cannot be 1), namely it is trivial. Hence, ${^*\!}\dot{S}^{\mu\nu}$ is also zero, and $p^\mu$ and $u^\mu$ are parallel, $p^\mu=mu^\mu$, which implies (cf. section \ref{our-condition}, but mind we do {\em not} assume $\dot{V}^\mu=0$ now)
\begin{align*}
  &m={\cal M}, \quad \mu=\gamma m, \quad \dot{s}=0, \\
  &\dot{s}^\mu=V^\mu\dot{V}_\nu s^\nu, \quad
   s^2\dot{V}^\mu=-s^\mu\dot{s}_\nu V^\nu.
\end{align*}

In a stationary situation, one can also assume
\begin{equation}
  \frac{\rm d}{{\rm d}\tau}(p_\mu s^\mu)=
  m\,\frac{\rm d}{{\rm d}\tau}(u_\mu s^\mu)=0 \,;
\end{equation}
in fact products between any dot-derivatives of $p^\mu$ (thus $u^\mu$) and $s^\mu$ should be constant as well, which brings the chain of relations
\begin{align*}
  \dot{p}_\mu s^\mu &= -p_\mu\dot{s}^\mu \,, \\
  \ddot{p}_\mu s^\mu &= -\dot{p}_\mu\dot{s}^\mu = p_\mu\ddot{s}^\mu \,, \\
  \dddot{p}_\mu s^\mu &= -\ddot{p}_\mu\dot{s}^\mu
  = \dot{p}_\mu\ddot{s}^\mu = -p_\mu\dddot{s}^\mu \, \\
  {\rm etc} \dots &
\end{align*}
Let us stress, on the other hand, that we do not a priori demand anything of scalars involving $V^\mu$, because this vector is auxiliary and need not respect the symmetry of the physical problem necessarily. However, in section \ref{our-condition-generalized} we saw that the $u^\mu\parallel p^\mu$ case offers one natural possibility: to prescribe $V^\mu(\tau)$ so that $s^\mu$ is -- and remains -- orthogonal to $p^\mu$.

\subsection{Other}

Other special properties are of course possible, though some of them are either contained in what has already been discussed, or they do not seem to lead to a particular simplification of the exercise.

For example, if $u_\alpha\dot{V}^\alpha=0$, then according to (\ref{pdotV=udotV}) $p_\alpha\dot{V}^\alpha=0$ as well, hence $\dot{S}_{\alpha\beta}\dot{V}^\beta=0$; since $\dot{V}^\beta$ is in general independent of the eigen-plane of $\dot{S}_{\alpha\beta}$ (section \ref{eigen-vectors}), $\dot{S}_{\alpha\beta}$ has to be trivial and we are back in situation mentioned in previous subsection.
Similarly, $u_\alpha\dot{s}^\alpha=0$ implies $p_\alpha\dot{s}^\alpha=0$ by (\ref{pdots=udots}), from where also $\dot{S}_{\alpha\beta}\dot{s}^\beta=0$ and again $\dot{S}_{\alpha\beta}=0$ generically.

An interesting case follows if $s_\alpha\dot{V}^\alpha=-\dot{s}_\alpha V^\alpha=0$. Then all the vectors $V^\mu$, $s^\mu$, $\dot{V}^\mu$, $\mu u^\mu-\gamma p^\mu$ are orthogonal to each other, so they form an orthogonal basis in space-time; this basis is actually the one we introduced in section \ref{ortho-basis}, just with the vectors $e^\mu_{\hat{2}}$ and $e^\mu_{\hat{3}}$ specially given by $\dot{V}^\mu$ and $\mu u^\mu-\gamma p^\mu$.

\section{Concluding remarks}

The spinning-particle problem is known to be inherently problematic, at least in the pole-dipole approximation, but it has a considerable history and brings a nice geometry. And it is not only of theoretical interest, as seen on growing interaction with the wide effort to understand and predict the generation of gravitational waves by collisions of compact objects, either by approximations or purely numerical approaches. In particular, an inspiral of a binary of compact bodies has become a key process in that field, and the role of spin in its outcome is a very lively topic -- see e.g. \cite{Ajith-11,BuonannoFH-13} or, especially for the question of the spin supplementary condition, \cite{PortoR-08}.
The effects of spin can mainly be expected to be important in final stages of the black-hole merger, because close to the horizon they are in fact stronger than radial attraction due to mass (in case of a single black hole, this is a defining property of the static limit, thus of the ergosphere), though with distance they fall much faster than the ``Newtonian" component. One particular situation where spin has been found to play a crucial role are the ``gravitational kicks" which the outcomes of binary black-hole mergers can get as a result of anisotropic emission of gravitational waves -- see e.g. \cite{HealyLZ-14} or, specifically for the role of the ``hidden momentum" in this effect, \cite{GrallaH-13}.

Let us mention, in particular, the extreme-mass-ratio limit of the binary inspiral, because it has been given a special attention recently, and because the spinning-particle problem represents its limit neglecting the radiation and approximating the small body by a test pole-dipole top \cite{SteinhoffP-12}. See, for example, \cite{BabakGC-15} for a review of this field, also including a discussion of different spin supplementary conditions. The extreme-mass-ratio instance of gravitational recoils has been studied e.g. by \cite{NagarHBZ-14}.

A decent history must of course be also attributed to the area of algebraic structure of curvature (Petrov types). Although in the neighbourhood of compact-body astrophysics it looks more academic, it is not fully so. Actually, the Kerr solution of Einstein's equations, celebrated by S. Chandrasekhar mainly as an {\em astrophysical} discovery, and indeed by default considered by astrophysicists when speaking of galactic nuclei or some binary X-ray sources, {\em is} algebraically special (type D). If the field in the core of these systems (of galactic nuclei in particular) is really close to the Kerr one, its special structure should reveal, for example, in the inspiral of some much-lower-mass compact body (mentioned above) and, consequently, also in the generated waves. In the monopole-test-particle limit, the ``Kerr-like" algebraic type is actually necessary for the geodesic equation (and also several other important problems) to be completely integrable \cite{Kubiznak-10}, hence not allowing for chaotic behaviour. However, even in type-D space-times the motion {\em is} in general chaotic if the particle is endowed with spin (\cite{Han-08} and references therein) or higher multipoles. The issues of spin, curvature structure and orbital dynamics are thus naturally bound together within one of today's major application directions of general relativity. We have not considered any gravitational waves emitted by the orbiting particle in this paper, but the background space-time certainly {\em can} contain waves. In this respect, the often treated Petrov type N is of physical relevance as a possible approximation of the far-zone radiation fields of bounded sources (see e.g. \cite{Bicak-00} and references therein).

To summarize the present paper, we have first reviewed the Mathisson--Papapetrou--Dixon (MPD) formulation, derived (or quoted) some useful relations of generic validity, while mainly focusing on the role of the spin supplementary condition. In the second part we projected the MPD equation of motion onto a suitable tetrad and expressed the spin-curvature term on its right-hand side in terms of the Weyl scalars $\Psi_{0\div 4}$ obtained in the complex null (Newman--Penrose) tetrad related to the orthonormal one. Specifically, we have chosen the orthonormal tetrad tied to the ``reference" observer $V^\mu$ fixing the spin condition, taking the corresponding spin vector $s^\mu$ (or rather its unit form) as one of spatial legs. In such a tetrad, the MPD equation appears as (\ref{V,dotp})--(\ref{e3,dotp}) which does not at all contain the null-tetrad scalars $\Psi_0$ and $\Psi_4$. The remaining two spatial vectors can be chosen in various ways, of which preferable are those fixing them ``intrinsically", along some directions provided by the geometry of the problem itself. We described one such possibility, applicable when $u^\mu\nparallel p^\mu$ and given by $(\mu u^\mu-\gamma p^\mu)$ and the ``vector product" of the former three which can be expressed as $(s^2\delta^\mu_\nu-s^\mu s_\nu)\,\dot{V}^\nu$.

Having expressed the MPD equation in terms of the Weyl scalars, it is natural to ask whether it assumes any special, simple form in space-times of particular Petrov types. For such a purpose, it is advantageous to choose $V^\mu$ so that the highest-multiplicity principal null direction of the Weyl tensor fall within the eigen-plane of the spin bivector, and to make it the first vector of the associated null tetrad. Even more favourable would be to make the null tetrad ``transverse" in the sense that the corresponding $\Psi_1$ and $\Psi_3$ projections vanish; the spinning-particle motion would then be fully determined by $\Psi_2$ and by the cosmological constant. Unfortunately, such a turn can only be reconciled with the spin structure in exceptional cases, namely, when it is possible -- by a suitable choice of $V^\mu$ -- to identify the spin eigen-plane with the real-vector plane of some of the transverse tetrads.

In the last part, we first treated where the exercise leads for the main spin conditions considered in the literature, revisited in particular the condition $\dot{V}^\mu=0$ (ensuring the very advantageous arrangement $u^\mu\parallel p^\mu$) and generalized it, suggesting also a natural resolution of non-uniqueness of the corresponding reference observer $V^\mu$. Then we checked how are the equations compatible with several particular types of motion.

Our next plan is to compare the analysis with that made in a different interpretation tetrad, namely the one tied to the word-line tangent $u^\mu$, and also to consider the case of massless particles.

\begin{acknowledgments}
We thank the grants GACR-14-10625S of the Czech Science Foundation (O. S.) and SVV-260211 of the Charles University (M. Š.) for support, L. F. O. Costa and T. Ledvinka for very helpful comments, and G. Lukes- Gerakopoulos for discussions.
\end{acknowledgments}

\appendix

\section{Orthonormal-tetrad and null-tetrad components of the Weyl tensor}
\label{Appendix-A}

Here we list the Weyl-tensor components in {\em some} orthonormal tetrad $(V^\mu,e^\mu_{\hat{\imath}})$ and in the related Newman--Penrose null tetrad $(k^\mu,l^\mu,m^\mu,\bar{m}^\mu)$. Note that we actually need Weyl-tensor {\em dual} in the Mathisson--Papapetrou--Dixon equations, but the null-tetrad scalars $\Psi_0$, \dots, $\Psi_4$ only change by an imaginary unit when dualizing $C_{\alpha\beta\gamma\delta}$, so it does not matter if we write the relations for $C_{\alpha\beta\gamma\delta}$ itself or for its dual. Note also that ${^*C}_{\alpha\beta\gamma\delta}=C^*_{\alpha\beta\gamma\delta}$ has the same symmetries as $C_{\alpha\beta\gamma\delta}$ (and hence 10 independent components).

Besides the ``automatic" properties following from the Riemann-tensor--type symmetries, the additional vanishing of the Weyl-tensor non-trivial trace, $g^{\alpha\gamma}C_{\alpha\beta\gamma\delta}=0$, implies another useful relations (when projected onto various null dyads),
\begin{eqnarray}
  C_{kmk\bar{m}}=C_{lml\bar{m}}=C_{mkml}=C_{\bar{m}k\bar{m}l}=0 \,, \\
  C_{km\bar{m}m}=\Psi_1, \quad C_{lm\bar{m}m}=\overline{\Psi}_3, \\
  C_{klkl}=C_{m\bar{m}m\bar{m}}=\Psi_2+\overline{\Psi}_2 \equiv 2\,{\rm Re}\Psi_2, \\
  C_{klm\bar{m}}=-\Psi_2+\overline{\Psi}_2 \equiv -2{\rm i}\,{\rm Im}\Psi_2,
\end{eqnarray}
where obvious notation has been used.
Direct substitution then yields the orthonormal components
\begin{align*}
  C_{\hat{0}\hat{1}\hat{0}\hat{1}}
      &= C_{klkl} = 2\,{\rm Re}\Psi_2\,, \\
  C_{\hat{0}\hat{2}\hat{0}\hat{2}}
      &= {\rm Re}\,C_{kml\bar{m}}+\frac{1}{2}\,{\rm Re}\,(C_{kmkm}+C_{lmlm}) \\
      &= -{\rm Re}\Psi_2+\frac{1}{2}\,({\rm Re}\Psi_0+{\rm Re}\Psi_4) \,, \\
  C_{\hat{0}\hat{3}\hat{0}\hat{3}}
      &= -C_{\hat{0}\hat{1}\hat{0}\hat{1}}-C_{\hat{0}\hat{2}\hat{0}\hat{2}} \\
      &= -{\rm Re}\Psi_2-\frac{1}{2}\,({\rm Re}\Psi_0+{\rm Re}\Psi_4) \,, \\
  C_{\hat{1}\hat{2}\hat{1}\hat{2}}
      &= -C_{\hat{0}\hat{3}\hat{0}\hat{3}} \,, \\
  C_{\hat{1}\hat{3}\hat{1}\hat{3}}
      &= -C_{\hat{0}\hat{2}\hat{0}\hat{2}} \,, \\
  C_{\hat{2}\hat{3}\hat{2}\hat{3}}
      &= -C_{\hat{0}\hat{1}\hat{0}\hat{1}} \,,
\end{align*}
\begin{align*}
  C_{\hat{0}\hat{1}\hat{0}\hat{2}} = C_{\hat{3}\hat{1}\hat{3}\hat{2}}
      &= {\rm Re}\,C_{klmk}+{\rm Re}\,C_{klml} \\
      &= {\rm Re}\Psi_3-{\rm Re}\Psi_1  \,, \\
  C_{\hat{0}\hat{1}\hat{0}\hat{3}} = C_{\hat{2}\hat{1}\hat{2}\hat{3}}
      &= -{\rm Im}\,C_{klkm}+{\rm Im}\,C_{klml} \\
      &= -{\rm Im}\Psi_1-{\rm Im}\Psi_3 \,, \\
  C_{\hat{0}\hat{2}\hat{0}\hat{3}} = C_{\hat{1}\hat{2}\hat{1}\hat{3}}
      &= \frac{1}{2}\,{\rm Im}\,(C_{kmkm}+C_{lmlm}) \\
      &= \frac{1}{2}\,({\rm Im}\Psi_0-{\rm Im}\Psi_4) \,, \\
  C_{\hat{0}\hat{1}\hat{2}\hat{1}} = -C_{\hat{0}\hat{3}\hat{2}\hat{3}}
      &= {\rm Re}\,C_{klkm}+{\rm Re}\,C_{klml} \\
      &= {\rm Re}\Psi_1+{\rm Re}\Psi_3 \,, \\
  C_{\hat{0}\hat{1}\hat{3}\hat{1}} = -C_{\hat{0}\hat{2}\hat{3}\hat{2}}
      &= {\rm Im}\,C_{klkm}+{\rm Im}\,C_{klml} \\
      &= {\rm Im}\Psi_1-{\rm Im}\Psi_3 \,, \\
  C_{\hat{0}\hat{2}\hat{1}\hat{2}} = -C_{\hat{0}\hat{3}\hat{1}\hat{3}}
      &= \frac{1}{2}\,{\rm Re}\,(C_{kmkm}-C_{lmlm}) \\
      &= \frac{1}{2}\,({\rm Re}\Psi_0-{\rm Re}\Psi_4) \,,
\end{align*}
\begin{align*}
  C_{\hat{0}\hat{1}\hat{2}\hat{3}}
      &= -{\rm i}\,C_{klm\bar{m}} = -2\,{\rm Im}\Psi_2 \,, \\
  C_{\hat{0}\hat{2}\hat{1}\hat{3}}
      &= {\rm Im}\,C_{kml\bar{m}}+\frac{1}{2}\,{\rm Im}\,(C_{kmkm}-C_{lmlm}) \\
      &= -{\rm Im}\Psi_2+\frac{1}{2}\,({\rm Im}\Psi_0+{\rm Im}\Psi_4) \,, \\
  C_{\hat{0}\hat{3}\hat{1}\hat{2}}
      &= C_{\hat{0}\hat{2}\hat{1}\hat{3}}-C_{\hat{0}\hat{1}\hat{2}\hat{3}} \\
      &= {\rm Im}\Psi_2+\frac{1}{2}\,({\rm Im}\Psi_0+{\rm Im}\Psi_4) \,.
\end{align*}
(Not all these are independent, needless to say. Others can be obtained just using the $C_{[\mu\nu][\kappa\lambda]}$ antisymmetries and the $C_{[\mu\nu]\leftrightarrow[\kappa\lambda]}$ symmetry.)
The respective components of the {\em dual} Weyl tensor are obtained according to
\[\Psi\rightarrow{^*}\Psi={\rm i}\Psi: \quad
  {\rm Re}({^*}\Psi)=-{\rm Im}\Psi, \quad {\rm Im}({^*}\Psi)={\rm Re}\Psi.\]

\subsection{Electric and magnetic curvature}
\label{Weyl-electric,magnetic}

Let us look into which Weyl scalars enter the Weyl-tensor electric and magnetic parts. These are introduced, in analogy with electric and magnetic parts of the Faraday tensor, as projections of the Weyl tensor on some time-like vector field (in our case represented by $V^\mu$),\footnote
{Alternatively, one could introduce these tensors by projections onto $u^\mu u^\nu$ or $p^\mu p^\nu/{\cal M}^2$, so ``as measured by the particle"; cf. \cite{CostaNZ-12} where the $u^\mu u^\nu$ projection is employed. In our case it is more natural to use $V^\mu V^\nu$, since we are using $V^\mu$ as the time vector of the interpretation basis.}
\begin{align}
  E_{\alpha\beta} &:= C_{\alpha\mu\beta\nu}V^\mu V^\nu
               \equiv C_{\alpha\hat{0}\beta\hat{0}} \,, \\
  B_{\alpha\beta} &:= {^*}C_{\alpha\mu\beta\nu}V^\mu V^\nu
               \equiv {^*}C_{\alpha\hat{0}\beta\hat{0}} \,.
\end{align}
Orthonormal components of this (symmetric) gravitoelectric and gravitomagnetic tidal fields can be seen above, namely, $E_{\hat{\imath}\hat{\jmath}}$ are given by real parts of $\Psi_a$, except $E_{\hat{1}\hat{3}}$ and $E_{\hat{2}\hat{3}}$ which are given by imaginary parts of $\Psi_1$ and $\Psi_3$, or of $\Psi_0$ and $\Psi_4$, respectively; in $B_{\hat{\imath}\hat{\jmath}}$, the appearance of real and imaginary parts is reversed.

In the so-called transverse orthonormal frames (section \ref{transverse}), the first MPD equation can in favourable cases be expressed in terms of just $\Psi_2$, which underlines the importance of type-D curvature where the latter is the only relevant Weyl scalar. If keeping only $\Psi_2$, we see that the above tidal fields have only diagonal components,
\begin{align*}
  & E_{\hat{1}\hat{1}}=-2E_{\hat{2}\hat{2}}=-2E_{\hat{3}\hat{3}}
    =2\,{\rm Re}\Psi_2 \,, \\
  & B_{\hat{1}\hat{1}}=-2B_{\hat{2}\hat{2}}=-2B_{\hat{3}\hat{3}}
    =-2\,{\rm Im}\Psi_2 \,.
\end{align*}

\subsection{Rotations within the $(m^\mu,\bar{m}^\mu)$ plane}
\label{m-rotations}

Of the well known 4 basic types of the NP-frame transformations, namely null rotations preserving $k^\mu$ or $l^\mu$, boosts in the $(k^\mu,l^\mu)$ plane and spatial rotations in the $(m^\mu,\bar{m}^\mu)$ plane, the last ones are mainly of interest for us, because in our problem $m^\mu$ and $\bar{m}^\mu$ can be rotated arbitrarily within the plane orthogonal to $V^\mu$ and $s^\mu$. Parametrizing such a rotation as
\begin{equation}
  m'^\mu=\exp({\rm i}\alpha)\,m^\mu, \quad
  \bar{m}'^\mu=\exp(-{\rm i}\alpha)\,\bar{m}^\mu,
\end{equation}
the Weyl scalars transform according to $\Psi_2'=\Psi_2$ and
\begin{eqnarray}
  \Psi_0'=\exp(2{\rm i}\alpha)\,\Psi_0, &\quad&
  \Psi_4'=\exp(-2{\rm i}\alpha)\,\Psi_4, \\
  \Psi_1'=\exp({\rm i}\alpha)\,\Psi_1, &\quad&
  \Psi_3'=\exp(-{\rm i}\alpha)\,\Psi_3
\end{eqnarray}
(hence those which were zero remain so).
In particular, $\Psi_1$ and $\Psi_3$ obey
\begin{eqnarray}
  {\rm Re}\Psi_1' &=& +{\rm Re}\Psi_1\,\cos\alpha-{\rm Im}\Psi_1\,\sin\alpha, \\
  {\rm Im}\Psi_1' &=& +{\rm Re}\Psi_1\,\sin\alpha+{\rm Im}\Psi_1\,\cos\alpha, \\
  {\rm Re}\Psi_3' &=& +{\rm Re}\Psi_3\,\cos\alpha+{\rm Im}\Psi_3\,\sin\alpha, \\
  {\rm Im}\Psi_3' &=& -{\rm Re}\Psi_3\,\sin\alpha+{\rm Im}\Psi_3\,\cos\alpha.
\end{eqnarray}
By such a rotation, it is generally possible to get rid of {\em one} component of {\em one} of these two scalars, but not more.

\bibliography{spinning.bib}

\end{document}